\documentclass[12pt]{article}

\usepackage[english]{babel} 
\usepackage[T1]{fontenc} 
\usepackage[utf8]{inputenc} 
\usepackage{graphicx}
\usepackage{amsmath}
\usepackage{amssymb}
\usepackage{amsthm}
\usepackage{slashed}
\usepackage{xcolor}
\usepackage{authblk}
\usepackage{array}
\usepackage{hyperref}

\usepackage[shortlabels]{enumitem}
\usepackage{ulem}   

\newcommand{\be}{\begin{equation}}
\newcommand{\ee}{\end{equation}}

\newcommand{\GeV}{\:\text{GeV}}
\newcommand{\TeV}{\:\text{TeV}}                                                 

\title{Revealing mass-degenerate states in Higgs boson signals}

\author[a]{Shehu AbdusSalam \thanks{abdussalam@sbu.ac.ir}}
\author[b]{Maria Eugenia Cabrera \thanks{meugenia@ecfm.usac.edu.gt}}

\affil[a]{Department of Physics, Shahid Beheshti University, Tehran
  19839, Islamic Republic of Iran.}
\affil[b]{Instituto de Investigaci\'on en Ciencias F\'isicas y Matem\'aticas (ICFM-USAC), Universidad de San Carlos de Guatemala, Guatemala.}

\date{}
\begin{document}

\maketitle

\begin{abstract}
  The observed Higgs boson signals to-date could be due to having two
  quasi-degenerate $125 \GeV$ scalar states in Nature. This kind of scenario
  tallies well with the predictions from the Next-to-Minimal Supersymmetric
  Standard Model (NMSSM). We have analysed the phenomenological NMSSM Higgs
  boson couplings and derived a parameterization of the signal strengths
  within the two quasi-degenerate framework.  With essentially two parameters,
  it is shown that the combined strengths of the two quasi-degenerate Higgs
  states in the leptonic (and b-quark) decay channels depart from the Standard
  Model values in the opposite direction to those in the vector boson
  channels.  We identify experimental measurements for distinguishing a single
  from a double Higgs scenarios. The proposed parameterization can be used for
  benchmarking studies towards establishing the status of quasi-degenerate
  Higgs scenarios.
\end{abstract}

\section{Introduction} 
Higgs boson discovery represents the beginning of a new epoch for
fundamental physics. The precise measurements of its couplings is an 
important aim for particle physics which could possibly give hint to 
physics beyond the Standard Model. With current data, the Higgs properties are compatible  
with the prediction of the Standard Model \cite{Khachatryan:2016vau,Aad:2015gba}. 
These same properties could also be due to the 
combination of effects arising from having two quasi-degenerate scalar states
around $125 \GeV$. Such a tantalizing possibility have been predicted by new
physics models such as the Next-to Minimal Supersymmetric Standard Model
(NMSSM). The impact of the Higgs properties and precision measurements on the 
NMSSM scenarios with two quasi-degenerate scalars will contribute towards
sharpening our understanding of the Higgs boson data and Nature -- it could be
that the data might have already contain some indications for new physics.\\

The current state of findings from the Large Hadron Collider (LHC), i.e. the
absence of direct signals of physics beyond the Standard Model (BSM), has been 
forecasted for the case of supersymmetry (SUSY) by pre-LHC global fits of
 models to data. For instance, as pointed out in 
\cite{Cabrera:2008tj,Cabrera:2009dm,AbdusSalam:2009qd} the large mass of the
Higgs was already an indication for heavy supersymmetric mass spectra. 
Within such models, phenomenological studies could be done via two main approaches, namely the
simplified models approach \cite{Alwall:2008ag, Alves:2011wf} 
and the phenomenological model parameterization 
\cite{Djouadi:1998di, AbdusSalam:2008uv, Berger:2008cq, AbdusSalam:2009qd, AbdusSalam:2012ir}.
In this article, the latter approach will be used. \\

Several groups have addressed mass-degenerate Higgs scenarios 
within the NMSSM. Refs.~\cite{Gunion:2012he,
  Munir:2013wka, Moretti:2015bua} have considered two quasi-degenerate
Higgs states for the real and complex NMSSM, with a mass difference large
enough to use the narrow width approximation. Ref.~\cite{Das:2017tob} has gone beyond
the narrow width approximation and showed
that interference effects can account for up to 40\% of total cross
sections. To be able to conclude that departures from SM prediction are a
consequence of the existence of more than one resonance
\cite{Grossman:2013pt, David:2014jla} have proposed statistical test based
on the analysis of a signal strength matrix, where all the channels are 
considered independent. A simplified version of their results agrees with
what was proposed previously in \cite{Gunion:2012he}. In this article,
we focus on the possibility of
having two mass-degenerate states with different coupling structures that when
combined mimic a single Higgs features. The main aim is to derive a set of NMSSM
parameters most relevant for quasi-degenerate Higgs studies vis-\'a-vise
collider data. For this, the NMSSM doublet-singlet mixings structure 
\cite{Carena:2015moc, King:2012tr, Das:2017tob} of the Higgs sector will be used.\\

In section~\ref{sec:gammagamma} we review the production and decay ratios of
the two lightest NMSSM CP-even Higgs states. We focus on the couplings
of these to vector bosons and heavy quarks. In section~\ref{sec:scan}
we perform a scan of the parameters of the NMSSM while imposing that the
the two lightest CP-even Higgs states reproduce the mass of the
standard Higgs measured by the LHC. We describe the allowed parameter space regions 
and relevant parameter correlations. In section~\ref{sec:2cpevenh}
the sample is then used together with analytical relations for the couplings
and signal strengths to show that the the quasi-degenerate Higgs properties
can be explained approximately by using just two free parameters. We also 
we show how the superposition of two quasi-degenerate Higgs around 125~GeV could be in
agreement with current experimental results. Finally in section~\ref{sec:signal} we
analyse the sample based on signal strength ratios that can discriminate between
the single versus double resonance scenarios.  

\section{Higgs couplings to fermions and vector bosons } 
\label{sec:gammagamma} 

Right after the discovery of the Higgs the search for signals of physics
beyond Standard Model in the production and decay of the Higgs became a
priority. A possible excess in the $\gamma\gamma$ channel motivated a lot of
work, some of them within the NMSSM framework~\cite{Ellwanger:2011aa,Gunion:2012gc,Gunion:2012he,King:2012tr,Gherghetta:2012gb}. In
particular King et. al.~\cite{King:2012tr} pointed out that the signal
strengths of the $\gamma\gamma$ channels could be enhanced for large
singlet-double mixing. We will take these 
as a starting point for analysing two quasi-degenerate 
CP-even Higgs states. \\

For the discussion of the following sections it is important to have a clear
picture of how the widths and therefore the Higgs branching ratios depend on
the singlet-doublet mixing. Let us start introducing some notation, we define
$\psi = (H_d,H_u,s)$ and $\phi = (h_0,H_0,s)$ in such a way that
$\langle h_0 \rangle=v$ and $\langle H_0 \rangle=0$:
\begin{eqnarray}
  \label{eq:notation}
   \phi_i=N_{ij}\psi_j 
\end{eqnarray}
where
\begin{eqnarray}
  \label{eq:Nmixmat}
  N = 
  \left( \begin{array}[h]{ccc} \cos\beta & \sin\beta & 0 \\ %
      \sin\beta & -\cos\beta & 0 \\ %
      0 & 0 & 1 \end{array} \right). 
\end{eqnarray}
The Higgs states $h=(h_1,h_2,h_3)$ are related to $\psi$ and $\phi$ in the
following way,
\begin{eqnarray}
  \label{eq:mixmatrix}
  h_i=U_{ij} \phi_j
\end{eqnarray}
where $U_{ij}$ are the elements of the mixing matrix, $U$.
We consider it convenient to use the elements of $U$ to parameterise
the couplings; for example $U_{i1}$ and $U_{i2}$ are respectively the $h_0$-component and
$H_0$-component of $h_i$. In this way it is easier to
make the comparison to the standard Higgs.\\

Using the above notation we write the tree-level Higgs
couplings to vector bosons and heavy quarks as:
\begin{eqnarray}
  \label{eq:couplings} \nonumber
  g_{h_iZZ} &=& g_{\mu\nu}\frac{g_1^2+g_2^2}{\sqrt{2}}\, v^2\,U_{i1}, \\ \nonumber
  g_{h_iWW} &=& g_{\mu\nu}\frac{g_2^2}{\sqrt{2}}\, v^2\, U_{i1},\\ \nonumber
  g_{h_itt} &=& \frac{m_t}{\sqrt{2}v} [\, U_{i1} -\cot\beta\, U_{i2}\,],\\
  g_{h_ibb} &=& \frac{m_b}{\sqrt{2}v} [\, U_{i1} +\tan\beta\, U_{i2}\,].
\end{eqnarray}
In the $H_0$ decoupling limit (i.e. $U_{12}=U_{22}=0$) all the couplings are
proportional to $U_{11}$, the $h_0$-component of $h_1$.  We are interested in
the departure of the production and decay signals of $h_1$ in the
$Z_3$-invariant NMSSM with respect to the one of the standard Higgs. To weight
this we will use the signal strength,
\begin{eqnarray}
  \label{eq:mufactordef}
  \mu = \frac{\sigma(SM\rightarrow h_i \rightarrow SM)|_{NMSSM}}{\sigma(SM\rightarrow h_{SM} \rightarrow SM)|_{SM}}
\end{eqnarray}
Because of the small width of the Higgs states we assume they are produced
on-shell, therefore the total cross sections are evaluated
as the production cross section times the branching ratio. \\

Now, in order to obtain the required properties for the Higgs states to
reproduce ATLAS and CMS measurements we consider two 
possibilities: %
\begin{enumerate}[I)]
\item $h_1$ or $h_2$ is the Higgs state detected at the LHC, and 
\item $h_1$ \textit{and} $h_2$ are the Higgs states measured by the LHC, where $h_1$ and
  $h_2$ are mass degenerate.
\end{enumerate}
We will show that these two possibilities correspond, respectively, to: %
\begin{enumerate}[I)]
\item Small singlet-doublet mixing, and 
\item Large singlet-doublet mixing.
\end{enumerate}%
Let us analyse the case with small singlet-doublet mixing where $h_1$ is
mainly $h_0$, in other words $U_{11}\sim 1$. For this case it is a good
approximation to consider that the width of $h_1$ is dominated by the decay
rate of $h_1\rightarrow b\bar{b}$ and therefore the variation of the width is
controlled by the square of the Higgs coupling to bottom quarks,
$g_{h_1b\bar{b}}$. Using the couplings described in eq.~(\ref{eq:couplings})
the signal strengths of the vector-boson fusion production of $h_1$ and
further decay to $WW/ZZ$ and $b\bar{b}$ are approximately, %
\begin{eqnarray}
  \label{eq:VBFzzww}
  \mu_{ \mathrm{VBF} \rightarrow h_1\rightarrow \mathrm{WW/ZZ}} %
  &\simeq& \hat{g}_{h_1WW}^2 \frac{\hat{g}_{h_1WW}^2}{\hat{g}_{h_1bb}^2} %
           = [ U_{11} ]^2 \ \frac{[ U_{11} ]^2}{[U_{11}+\tan\beta\,U_{12}]^2},\\
  \label{eq:VBFbb}
    \mu_{\mathrm{VBF} \rightarrow h_1\rightarrow b\bar{b}} %
  &\simeq& \hat{g}_{h_1WW}^2 \frac{\hat{g}_{h_1bb}^2}{\hat{g}_{h_1bb}^2} %
  = [ U_{11} ]^2,
\end{eqnarray}  
where $\hat{g} = g_{\mathrm{NMSSM}}/g_{\mathrm{SM}}$, the
couplings $g_{\mathrm{NMSSM}}$ are those in
eq.~(\ref{eq:couplings}), and $g_{\mathrm{SM}}$ are the Standard Model (SM) couplings. %
The enhancement or suppression of the first signal strengths depends on $\tan\beta\,\,U_{12}/U_{11}$.
As such, the absolute value and sign of this factor determines respectively the magnitude
of the ratio between the signal strengths and whether there is an enhancement or suppression of 
$\mu_{\mathrm{VBF} \rightarrow h_1\rightarrow \mathrm{WW/ZZ}}$ with respect to
$\mu_{\mathrm{VBF} \rightarrow h_1\rightarrow b\bar{b}}$. A 
similar analysis holds when $h_2$ is considered the Higgs state measured at the LHC. \\

Next, let us examine the case with large singlet-doublet mixing where $h_1$
has non-negligible S content. 
In this case, the approximation $U_{11}\sim 1$ is not valid any more.
The assumption that the width of $h_1$ is almost totally controlled by 
$h_1\rightarrow b\bar{b}$ is no longer a good approximation. 
The size of $\tan\beta\,U_{12}/U_{11}$ may take very large values and 
therefore the branching ratio could significantly differ with respect to 
the standard Higgs. 
So, we would like to have a simple expression for the widths appropriate for all
values of $U_{i1}$. In terms of the standard Higgs decay rates, one can write 
\begin{eqnarray}
  \label{eq:width0}
  \Gamma_i &=& \Gamma_{h_i\rightarrow bb/\tau\tau} + \Gamma_{h_i\rightarrow
               WW/ZZ} + \Gamma_{h_i\rightarrow SM_\mathrm{rest}} \\
           &=& \hat{g}_{h_i bb/\tau\tau}^2\,\Gamma_{(h_{SM}\rightarrow bb/\tau\tau)}\,
               + \hat{g}_{h_i WW/ZZ}^2\,\Gamma_{(h_{SM}\rightarrow WW/ZZ)}\, 
               +(U_{i1})^2\,\Gamma_{(h_{SM}\rightarrow SM_\mathrm{rest})}
\end{eqnarray}
where $h_i\rightarrow \mathrm{SM}_{\mathrm{rest}}$ represents the rest of the
decay channels. The dominant contribution for the rest of decay channels is
the decay to gluons through a top loop. 
For simplicity we are going to consider that the rest of the decay modes
behave as the ones of the standard Higgs. For this reason we took the corresponding
decay rate proportional to the square of $h_i$'s $h_0$ content, $U_{i1}$. By writing
the decay rates in terms of the SM branching ratios we get
\begin{eqnarray}
  \label{eq:width}
  \Gamma_i/\Gamma_{\mathrm{SM}} &\simeq& \mathrm{BR}_{h_{SM}\rightarrow bb/\tau\tau}\,
                    \hat{g}_{h_i bb/\tau\tau}^2 +
                    \mathrm{BR}_{h_{SM}\rightarrow WW/ZZ}\, 
                    \hat{g}_{h_i WW/ZZ}^2 +
                    \mathrm{BR}_{h_{SM}\rightarrow SM}\, (U_{i1})^2\\ 
  &\simeq& \mathrm{BR}_{h_{SM}\rightarrow bb/\tau\tau}
                                (U_{i1}+U_{i2}\tan\beta)^2
                                + (1-\mathrm{BR}_{h_{SM}\rightarrow bb/\tau\tau})
                                (U_{i1})^2. 
\end{eqnarray}
For large singlet-doublet mixing the widths of $h_1$ and $h_2$ could be much
smaller than $\Gamma_{\mathrm{SM}}$, producing large departures of the
branching ratios with respect to the ones of the standard Higgs, unless
  the widths and the decay rates of each Higgs state change at the same
  proportion. From now on we will use eq.~(\ref{eq:width}) as the
  enhancement(suppression) rate of the width
  with respect to the SM value.\\

The analytic expressions for the signal strengths for vector-boson fusion
production and decay to $WW/ZZ$ and $b\bar{b}$ can be written as,
\begin{eqnarray}
  \label{eq:mutot1}
  \mu_{VBF\rightarrow h_1 \rightarrow WW/ZZ}^{\mathrm{an}} &\simeq&
                            \frac{(U_{11})^4}{(1-\mathrm{BR}_{h_{SM}\rightarrow bb/\tau\tau})(U_{11})^2 +
                            \mathrm{BR}_{h_{SM}\rightarrow bb/\tau\tau}
                            (U_{11}+U_{12}\tan\beta)^2}, 
  \\ \label{eq:mutot2}
  \mu_{VBF\rightarrow h_2 \rightarrow WW/ZZ}^{\mathrm{an}} &\simeq&
                            \frac{(U_{21})^4}{(1-\mathrm{BR}_{h_{SM}\rightarrow bb/\tau\tau})(U_{21})^2 +
                            \mathrm{BR}_{h_{SM}\rightarrow bb/\tau\tau}
                            (U_{21}+U_{22}\tan\beta)^2},
  \\ \label{eq:mutot3}
  \mu_{VBF\rightarrow h_1 \rightarrow bb}^{\mathrm{an}} &\simeq&
                            \frac{(U_{11})^2 (U_{11}+\tan\beta U_{12})^2}
                            {(1-\mathrm{BR}_{h_{SM}\rightarrow bb/\tau\tau})(U_{11})^2 +
                            \mathrm{BR}_{h_{SM}\rightarrow bb/\tau\tau}
                            (U_{11}+U_{12}\tan\beta)^2},
  \\  \text{ and } \label{eq:mutot4}
  \mu_{VBF\rightarrow h_2 \rightarrow bb}^{\mathrm{an}} &\simeq&
                            \frac{(U_{21})^2 (U_{21}+\tan\beta U_{22})^2}
                            {(1-\mathrm{BR}_{h_{SM}\rightarrow bb/\tau\tau})(U_{21})^2 +
                            \mathrm{BR}_{h_{SM}\rightarrow bb/\tau\tau}
                            (U_{21}+U_{22}\tan\beta)^2}.
\end{eqnarray}
Note that for a large singlet-doublet mixing the relative size of
$\tan\beta\,U_{12}/U_{11}$ has a larger range of variation than in the case of
small singlet-doublet mixing, as consequence there might be larger
enhancement(suppression) to the signals.  Moreover, since the $H_0$-component
of the Higgs states is the one responsible for large variations of the
branching ratios, it is interesting to see that in the $H_0$ decoupling limit
($U_{12}\simeq 0$ and $U_{22}\simeq 0$),
\begin{displaymath}
     \lim_{m_{H_0}\gg\, m_{h_0},m_S} \mu_{VBF\rightarrow h_i \rightarrow WW/ZZ}^{\mathrm{an}} \simeq
                            (U_{i1})^2\ \ , \textrm{ and } \qquad %
   \lim_{m_{H_0}\gg\, m_{h_0},m_S} \mu_{VBF\rightarrow h_i \rightarrow bb}^{\mathrm{an}} \simeq
                            (U_{i1})^2.
\end{displaymath}
Hence for large singlet-doublet mixing it is not possible to reproduce the
experimental data with a single Higgs state. But, if $h_1$ and $h_2$ are
mass quasi-degenerate, assumed to be unresolved away from each other by experiments, the
superposition of the two states could show up in signals as single standard Higgs with,
\begin{eqnarray}
  \label{eq:largemH0}
  && \lim_{m_{H_0}\gg m_{h_0},m_S} \mu_{VBF\rightarrow h_1 \rightarrow
  WW/ZZ}^{\mathrm{an}} + \mu_{VBF\rightarrow h_2 \rightarrow
  WW/ZZ}^{\mathrm{an}} \simeq (U_{11})^2 + (U_{21})^2 \sim 1 \textrm{ and } \\ 
  && \lim_{m_{H_0}\gg m_{h_0},m_S} \mu_{VBF\rightarrow h_{1} \rightarrow
  bb}^{\mathrm{an}} + \mu_{VBF\rightarrow h_{2} \rightarrow
  bb}^{\mathrm{an}} \simeq  (U_{11})^2 + (U_{21})^2 \sim 1.  
\end{eqnarray}
Notice that the last (approximate)equalities require $U_{31}\simeq 0$ to fulfill the unitarity
condition for U.\\

It is interesting to compare the departure of the signal strengths for different
channels of the same Higgs state. As described earlier, the ratio between
signal strengths depends on $\tan\beta\, U_{12}/U_{11}$ for $h_1$ and on $\tan\beta\,U_{22}/U_{21}$ for $h_2$.
As such, the departure of the global signal strength will depend on the relation between $U_{12}$
and $U_{22}$.\\

In the following sections we analyse the scenario with large singlet-doublet
mixing. We will assume that the Higgs signal measured by ATLAS and CMS is a
superposition of the production and decay of two Higgs states.  To get the
global enhancement(suppression) we will sum the contribution of the two Higgs
states. Notice that for this approximation to be valid the widths should be
much smaller that the mass difference between $h_2$ and $h_1$.

\section{The phenomenological NMSSM Parameters scan}
\label{sec:scan}

Let us consider the case where the Higgs signal measured by ATLAS and CMS is a
superposition of the production and decay of $h_1$ and $h_2$, meaning that the
Higgs states are close enough not to be resolved by the experiments, but with
large enough separation to have negligible interference effects. To study the
region of the parameter space of the NMSSM where this condition is fulfilled
we perform a parameter scan as done in \cite{AbdusSalam:2017uzr}.

\subsection{The phenomenological NMSSM (pNMSSM)} 
We shall consider an R-parity conserving NMSSM with superpotential, 
\be \label{superpot}
W_{NMSSM} = W_{MSSM'} - \epsilon_{ab}\lambda {S} {H}^a_1 {H}^b_2 + \frac{1}{3}
\kappa {S}^3 \ ,
\ee 
where 
\begin{eqnarray}
  W_{MSSM'}&=& \epsilon_{ab} \left[ 
    (Y_E)_{ij} H_1^a    L_i^b    {\bar E}_j 
    + (Y_D)_{ij} H_1^a    Q_i^{b}  {\bar D}_{j} 
    + (Y_U)_{ij} H_2^b    Q_i^{a}  {\bar U}_{j}
    \right].
  \label{wmssm}
\end{eqnarray}
The chiral superfields have the following $SU(3)_C\otimes SU(2)_L\otimes
U(1)_Y$ quantum numbers, 
\begin{eqnarray} 
  && L:(1,2,-\frac{1}{2}),\quad {\bar E}:(1,1,1),\\
  && Q:\,(3,2,\frac16),\quad
  {\bar U}:\,(\bar{3},1,-\frac{2}{3}),\quad {\bar D}:(\bar{3},1,\frac13), \\
  && H_1:(1,2,-\frac{1}{2}),\quad  H_2:\,(1,2,\frac{1}{2}). 
  \label{fields}
\end{eqnarray}
The corresponding soft SUSY-breaking terms are
\be
V_\mathrm{soft} = V_2 + V_3 + m_\mathrm{S}^2 | S |^2 +
(-\epsilon_{ab}\lambda A_\lambda {S} {H}^a_1 {H}^b_2 + 
\frac{1}{3} \kappa A_\kappa {S}^3  
+ \mathrm{H.c.}), \ee
with 
\begin{eqnarray}
  V_2 & = & m_{H_1}^2 {{H^*_1}_a} {H_1^a} + m_{H_2}^2 {{H^*_2}_a}
  {H_2^a} + 
  {\tilde{Q}^*}_{i_La} (m_{\tilde Q}^2)_{ij} \tilde{Q}_{j_L}^{a} +
  {\tilde{L}^*}_{i_La} (m_{\tilde L}^2)_{ij} \tilde{L}_{j_L}^{a}  
  + \nonumber \\ &&
  \tilde{u}_{i_R} (m_{\tilde u}^2)_{ij} {\tilde{u}^*}_{j_R} +
  \tilde{d}_{i_R} (m_{\tilde d}^2)_{ij} {\tilde{d}^*}_{j_R} +
  \tilde{e}_{i_R} (m_{\tilde e}^2)_{ij} {\tilde{e}^*}_{j_R}, \\
  V_3 & = & \epsilon_{ab} \sum_{ij}
  \left[
    (T_E)_{ij} H_1^a \tilde{L}_{i_L}^{b} \tilde{e}_{j_R}^* +
    (T_D)_{ij} H_1^a \tilde{Q}_{i_L}^{b}  \tilde{d}_{j_R}^* +
    (T_U)_{ij}  H_2^b \tilde{Q}_{i_L}^{a} \tilde{u}_{j_R}^*
    \right]
  + \mathrm{H.c.}. 
\end{eqnarray}
A tilde-sign over the superfield symbol represents the scalar
component. However, an asterisk over the superfields as in, for example, 
$\tilde{u}_R^*$ represents the scalar component of $\bar{U}$. 
The $SU(2)_L$ fundamental representation indices are donated by
$a,b=1,2$ while the generation indices by
$i,j=1,2,3$. $\epsilon_{12}=\epsilon^{12}=1$ is a totally
antisymmetric tensor.

In an approach similar to that of the pMSSM~\cite{Djouadi:1998di,
  AbdusSalam:2008uv, Berger:2008cq, AbdusSalam:2009qd}, the pNMSSM parameters
are defined at the weak scale with the non-Higgs sector set,
\be \label{susypart} M_{1,2,3};\;\; m^{3rd \,
  gen}_{\tilde{f}_{Q,U,D,L,E}},\;\; m^{1st/2nd \,
  gen}_{\tilde{f}_{Q,U,D,L,E}}; \;\;A_{t,b,\tau}.  \ee
Here, $M_{1,2,3}$ and
$m_{\tilde f}$ are respectively the gaugino and the sfermion mass
parameters. $A_{t,b,\tau}$ represent the trilinear scalar couplings.
With electroweak symmetry breaking,an effective $\mu$-term,
$\mu_\mathrm{eff} = \lambda \, v_s$ is developed. The $\mu$-term, the ratio of the MSSM-like Higgs
doublets' vevs $\tan \beta=\left<H_2\right>/\left<H_1\right>$ and the Z-boson mass, $m_Z$ lead to
the tree-level Higgs sector parameters
\be \label{Higgspars}
\tan\beta, \lambda, \kappa, A_{\lambda}, A_{\kappa}, \lambda \, v_s.  \ee
Next, including four SM nuisance parameters, namely, the top and bottom
quarks $m_{t,b}$, $m_Z$ and the strong coupling constant, $\alpha_s$, makes
the pNMSSM parameters: \be \label{theparams} \theta = \{
M_{1,2,3};\;\; m^{3rd \, gen}_{\tilde{f}_{Q,U,D,L,E}},\;\; m^{1st/2nd \,
  gen}_{\tilde{f}_{Q,U,D,L,E}}; \;\;A_{t,b,\tau,\lambda, \kappa}; \;\; \tan
\beta, \lambda, \kappa, \mu_\mathrm{eff}; \;\; m_{t,Z,b}, \alpha_s \}.  \ee

\subsection{The scanning procedure}
$M_{1,2}$ affects the gaugino masses for which a wide range, $\mathcal{O}(\mathrm{GeV})$ to $\mathcal{O}(\mathrm{TeV})$,
is possible. We let $M_1 \in [-4, 4] \TeV$ and same for $M_2 >0$. With the LHC in mind, 
we let the gluino and squark mass parameters be within $[100 \GeV, 4 \TeV]$, and the trilinear scalar couplings
allowed in $[-8 \TeV, 8 \TeV]$. $\tan \beta$ is allowed between 2 and
60. For minimising fine-tuning, we subjectively let $\mu_\mathrm{eff} = \lambda \, v_s$ to vary within
100 to 400 GeV not too far away from the Z-boson mass. The remaining Higgs-sector
parameters were set within the ranges shown in Table~\ref{tab.param}.
\begin{table}[htbp!] 
  \begin{center}{\begin{tabular}{|lll|}
        \hline
        Parameter & Range       & Posterior range \\ 
        \hline
        $M_1$                   & [$-4$ TeV, $4$ TeV]   &    \\
        $M_2$                   & [$0$ TeV, $4$ TeV]    &   \\
        $M_3, \,\, m^{3rd \, gen, \,\, 1st/2nd \, gen}_{\tilde{f}_{Q,U,D,L,E}}$   & [$100$ GeV, $4$ TeV]  & \\   
        $A_{t,b,\tau}$ &  [-$8$ TeV, $8$ TeV]  & \\ 
        $\tan \beta$ &  [$2$, $60$]            & [$8.8$, $28.3$]\\ 
        $\lambda$ & [$10^{-4}$, $0.75$]         & [$0.17$, $0.52$]\\
        $\kappa$ & [$-0.75$, $0.75$]           & [$-0.50$, $0.75$]\\
        $\mu_\mathrm{eff}$ & [$100$, $400$] GeV  & [$111$, $308$] GeV\\
        $A_{\lambda}$ &  [$50$ GeV, $4$ TeV]     & [$1.34$, $4$] TeV\\ 
        $A_{\kappa}$ &  [$-2$ TeV, $2$ TeV]      & [$-1646$, $846$] GeV\\ 
        \hline
        $m_t$  &  172.6 $\pm$ 1.4 GeV  & \\
        $m_Z$  &  91.1876 $\pm$ 0.0021 GeV  & \\
        $m_b(m_b)^{\overline{MS}}$  &  4.20 $\pm$ 0.07 GeV  & \\
        $\alpha_{s}(m_Z)^{\overline{MS}}$ & 0.1172 $\pm$ 0.002   & \\ 
        \hline
  \end{tabular}}\end{center}
\caption{The 26 pNMSSM parameters and their corresponding flat prior
  probability density distribution ranges. The SM parameters were varied
  according to Gaussian distributions with the shown central values and
  standard deviations.  The third column (to be addressed in section.~\ref{sec:2cpevenh}) shows the 95\% Bayesian
  confidence regions for the posterior sample used in
  Figure.~\ref{fig:mucomparison}. For this posterior sample,
  $m_{h_2}-m_{h_1}<3$~GeV with both $m_{h_1}$ and $m_{h_2}$ allowed within
  [$122$, $128$] GeV. \label{tab.param}}
\end{table}

The selected pNMSSM points were required pass all the constraints summarised in Tab.\ref{tab.obs}.
These are: the Higgs boson mass $m_h$, the neutralino cold dark matter
(CDM) relic density $\Omega_{CDM} h^2$, anomalous magnetic moment of the muon
$\delta a_\mu$, and the B-physics related limits summarised in the upper part
of Table~\ref{tab.obs}. The experimental constraints used were those
implemented in {\sc NMSSMTools} \cite{Ellwanger:2006rn, Ellwanger:2005dv,
  Djouadi:1997yw, Degrassi:2009yq, Domingo:2007dx, Domingo:2015wyn}, {\sc
  Lilith} \cite{Bernon:2015hsa}, {\sc MicrOMEGAs} \cite{Boos:2004kh,
  Semenov:2008jy, Belanger:2010st, Pukhov:1999gg, Belyaev:2012qa,
  Belanger:2013oya, Belanger:2010pz, Belanger:2008sj, Belanger:2006is,
  Barducci:2016pcb}, {\sc SModelS}'\cite{Ambrogi:2017neo,Kraml:2013mwa,
  Buckley:2013jua, Sjostrand:2006za, Beenakker:1996ch, Beenakker:1997ut,
  Kulesza:2008jb, Kulesza:2009kq, Beenakker:2009ha, Beenakker:2010nq,
  Beenakker:2011fu} implementation of ATLAS and CMS
limits\cite{ATLAS:2013cma, TheATLAScollaboration:2013fha, TheATLAScollaboration:2013lha,
  TheATLAScollaboration:2013tha, CMS:2014nia, CMS:2015kza,
  Chatrchyan:2013mys, Khachatryan:2014qwa, Chatrchyan:2014lfa,
  Khachatryan:2015vra, Chatrchyan:2013xna}, and {\sc HiggsBounds}
\cite{Bechtle:2011sb, Bechtle:2013wla, Aad:2014fia, Aad:2015xja,
  Aad:2014iia, Chatrchyan:2014tja, Aad:2014xva, Khachatryan:2015cwa,
  Aad:2014vgg, Aad:2015kna, Aad:2014yja, Khachatryan:2014jba,Sirunyan:2017exp,
  Khachatryan:2014ira, Aad:2015agg}. The Higgs boson signal strength
measurements from Tevatron \cite{Aaltonen:2013ioz}, ATLAS \cite{Aad:2015gba,
  Aad:2014eha, Aad:2015ona, Aad:2014eva, Aad:2015vsa, Aad:2015iha,
  Aad:2015gra, Aad:2014xzb, Aad:2014xva, ATLAS:2015yda, Aad:2014iia} and
CMS \cite{Khachatryan:2014jba,Sirunyan:2017exp, Khachatryan:2014ira, Chatrchyan:2013iaa,
  Chatrchyan:2013mxa, Chatrchyan:2014nva, Chatrchyan:2013zna,
  Khachatryan:2014qaa, Khachatryan:2015ila, Khachatryan:2015bnx,
  Chatrchyan:2014tja} as implemented in {\sc Lilith} v1.1 (with data version
15.09) \cite{Bernon:2015hsa} were also included.

\begin{table}
  \begin{center}{\begin{tabular}{|l l p{9.5cm}|}
        \hline
        Observable & Constraint   & References\\ 
        \hline
        $m_h$& $125.09 \pm 3.0$ GeV & \cite{Aad:2015zhl} \\
        $Br(B \rightarrow X_s \gamma)$ & $(3.32 \pm 0.16) \times
        10^{4}$ & \cite{Bobeth:1999ww, Buras:2002tp, Amhis:2014hma}\\  
        $Br(B_s \rightarrow \mu^+ \mu^-)$ & $(3.0 \pm 0.6) \times
        10^{-9}$ & \cite{Aaij:2017vad, Bobeth:2013uxa, Buras:2002vd}  \\ 
        $\Delta M_{B_s}$ & $17.757 \pm 0.021$ & \cite{Buras:2002vd,Ball:2006xx}\\
        $\Delta M_{B_d}$ & $0.5064 \pm 0.0019$ &   \cite{Buras:2002vd,Ball:2006xx}\\ 
        $Br(B_u \rightarrow \tau \nu)$ & $1.06 \pm 0.19$ &
        \cite{Barate:2000rc, Aubert:2004kz, Gray:2005ad, Akeroyd:2003zr}\\
        $\delta a_{\mu}$ & $(30.2 \pm 8.80) \times 10^{-10}$ &
        \cite{Bennett:2006fi, Domingo:2007dx, Domingo:2015wyn} \\        
        $\Omega_{CDM} h^2$ & $0.12 \pm 0.02$ & \cite{Ade:2015xua} \\
        \multicolumn{2}{|l}{Higgs signal strengths} & 
        \cite{Aaltonen:2013ioz, Aad:2015gba, Aad:2014eha,
          Aad:2015ona, Aad:2014eva, Aad:2015vsa, Aad:2015iha, Aad:2015gra,
          Aad:2014xzb, Aad:2014xva, ATLAS:2015yda, Aad:2014iia,
          Khachatryan:2014jba,Sirunyan:2017exp, Khachatryan:2014ira, Chatrchyan:2013iaa,
          Chatrchyan:2013mxa, Chatrchyan:2014nva, Chatrchyan:2013zna,
          Khachatryan:2014qaa, Khachatryan:2015ila, Khachatryan:2015bnx,
          Chatrchyan:2014tja} \\
        \multicolumn{2}{|l}{CDM direct detection limits} &  
        \cite{Akerib:2016vxi, Aprile:2017iyp, Tan:2016zwf, Amole:2015pla, Amole:2016pye, Akerib:2016lao, Fu:2016ega}\\
        \multicolumn{2}{|l}{Constraints in {\sc HiggsBounds}} & \cite{Bechtle:2011sb, Bechtle:2013wla,
          Aad:2014fia, Aad:2015xja, Aad:2014iia, Chatrchyan:2014tja, Aad:2014xva,
          Khachatryan:2015cwa, Aad:2014vgg, Aad:2015kna, Aad:2014yja,
          Khachatryan:2014jba, Sirunyan:2017exp, Khachatryan:2014ira, Aad:2015agg}\\
        \multicolumn{2}{|l}{Constraints in {\sc SModelS}} & \cite{Ambrogi:2017neo,Kraml:2013mwa,
          Buckley:2013jua, Sjostrand:2006za, Beenakker:1996ch, Beenakker:1997ut,
          Kulesza:2008jb, Kulesza:2009kq, Beenakker:2009ha, Beenakker:2010nq, Beenakker:2011fu, 
          ATLAS:2013cma, TheATLAScollaboration:2013fha, TheATLAScollaboration:2013lha, TheATLAScollaboration:2013tha, CMS:2014nia, CMS:2015kza,
          Chatrchyan:2013mys, Khachatryan:2014qwa, Chatrchyan:2014lfa, Khachatryan:2015vra, Chatrchyan:2013xna}\\
        \hline 
      \end{tabular}}\end{center}
  \caption{Summary of the central values and errors for the
    observables. Theoretical uncertainties have been added in quadrature
    to the experimental uncertainties quoted.}
  \label{tab.obs}
\end{table}

\subsection{Constraints on the parameters of the Higgs
  sector} \label{higgsconsts} From the pNMSSM parameter scan, we use a sample
with two quasi-degenerate lightest CP-even Higgs bosons.  It was required that
$h_1$ and $h_2$ have mass equal to $125\pm 3$~GeV, where the $\pm 3$~GeV
accounts to the theoretical errors associated to the values of the masses
computed by NMSSMtools. In addition it was required that the mass difference,
$m_{h_2}-m_{h_1}<3$~GeV \footnote{ The CMS resolutions for Higgs bosons are
  channel dependent and typically around 2.5 to
  4~GeV~\cite{Khachatryan:2014jba,Sirunyan:2017exp} for bosonic channels. As
  such $m_{h_2}-m_{h_1}<3$~GeV can be considered as a mass degeneracy
  condition for which the two Higgs cannot be resolved by CMS run-2.}.  We
focus on the regions of the Higgs sector parameters for studying the
correlations within those parameters and for relating them to other parameters
which are directly connected
with the signals measured at the LHC such as the CP-even Higgs mixing matrices. \\

It is useful to have an explicit form for the Higgs mixing matrix $U$. We
 parameterise this using three angles $\theta_{13}$, $\theta_{12}$, and 
$\theta_{23}$ such that 
\begin{eqnarray} \nonumber
  \label{eq:Umixmat2H}
  U &=& \left( \begin{array}[h]{ccc} c_{13} & 0 & s_{13} \\ %
                 -s_{13} & 0 & c_{13} \\ %
                 0 & 1 & 0 \end{array} \right)%
        \left( \begin{array}[h]{ccc} 1 & 0 & 0 \\ %
                 0 & c_{23} & s_{23} \\ %
                 0 & -s_{23} & c_{23} \end{array} \right ) %
        \left( \begin{array}[h]{ccc} c_{12} & s_{12} & 0 \\ %
                 -s_{12} & c_{12} & 0\\
                 0 & 0 & 1 \end{array} \right) \\ \nonumber
  \\
    &=&   \left( \begin{array}[h]{ccc} %
                   c_{13}c_{12}+s_{13}s_{23}s_{12} & c_{13}s_{12}-s_{13}s_{23}c_{12} & s_{13}c_{23} \\ %
                   -s_{13}c_{12}+c_{13}s_{23}s_{12} & -s_{13}s_{12}-c_{13}s_{23}c_{12} & c_{13}c_{23} \\ %
                   -c_{23}s_{12} & c_{23}c_{12} & s_{23} \end{array} \right ). 
\end{eqnarray}
Here $c_{ij}=\cos\theta_{ij}$ and $s_{ij}=\sin\theta_{ij}$. Given the mixing matrix,
obtained numerically by the SUSY spectra calculator NMSSMtools, then the mixing angles
can be extracted as: 
\begin{eqnarray}
  \label{eq:angles}
  s_{23}=U_{33}, \quad && \quad s_{13}=\frac{U_{13}}{c_{23}},\\
  s_{12}=-\frac{U_{31}}{c_{23}},
  \quad && \quad c_{13}=\frac{U_{23}}{c_{23}}.
\end{eqnarray}
Now, considering that we want to reproduce a standard Higgs signal, we
determine the expected ranges for the mixing angles. In order to get the ratio
between $\mu_{VBF\rightarrow h_{1,2} \rightarrow WW/ZZ}$ and
$\mu_{VBF \rightarrow h_{1,2} \rightarrow bb}$ close to one, either the value
of
$\mu_{VBF\rightarrow h_i \rightarrow WW/ZZ}/\mu_{VBF \rightarrow h_i
  \rightarrow bb}$ for each Higgs state has to be close to one, or a fine
cancellation should take place. In this work we focus on the first
case\footnote{In other words, this means that we restrict our analyses to the
  scenario where $H_0$ is much heavier than $h_0$ and $S$.}. From
eqs.~(\ref{eq:mutot1})-(\ref{eq:mutot4}) one can see that this condition is
possible when $U_{12}$ and $U_{22}$ are very small and as a result $s_{12}$
and $s_{23}$ should also be very small according to eq.~(\ref{eq:Umixmat2H}).
On an other hand, eq.~(\ref{eq:largemH0}) implies that the superposition of
$h_1$ and $h_2$ can reproduce the standard Higgs signal for $U_{31}\sim 0$
(i.e. large values of $m_{H_0}$). For this to happen either $\theta_{12}$ has
to be very small or $\theta_{23}$ has to be close to $\pm\pi/2$. In
  summary, $\theta_{12}\sim 0$ and $\theta_{23}\sim 0$ will guarantee that
  we are working in the regime where the superposition of the two Higgs states agrees with experimental measurements.\\

In the limit of small $\theta_{12}$ and $\theta_{23}$, 
\begin{displaymath}
  s_{12}\simeq \theta_{12}, \quad s_{23}\simeq \theta_{23}, %
  \quad c_{12}\simeq 1, \quad c_{23}\simeq 1
\end{displaymath}
and the mixing matrix eq.~(\ref{eq:Umixmat2H}) reduces to 
\begin{eqnarray}
  \label{eq:Umixmat2haprox}
    U &\simeq& \left( \begin{array}[h]{ccc} %
                   c_{13} & c_{13}\theta_{12}-s_{13}\theta_{23} & s_{13} \\ %
                    -s_{13} & -s_{13}\theta_{12}-c_{13}\theta_{23} & c_{13} \\ %
                   -\theta_{12} & 1 & \theta_{23} \end{array} \right )
\end{eqnarray}
where we have neglected $\mathcal{O}(\theta^2)$ terms. For the results of our
scan this approximation works with a $0.5\%$ error.\\ 

We have been able to constrain the parameters of the mixing matrix requiring
conditions that will give us a standard-like Higgs signal. This conditions will
affect the masses or couplings of the heaviest and pseudoscalar Higgs bosons. To see this, 
it will be useful to relate the
mixing angles $\theta_{13}$, $\theta_{23}$ and
$\theta_{12}$ to the fundamental parameters of the Higgs sector. 
Using eq.~(\ref{eq:Umixmat2H}) we relate the terms of the mass matrix with the
physical masses by introducing two new parameters: $m_h^2$, the central value of
the two lightest CP-even Higgs states, and $\delta m_h^2$, half of the squared
mass difference,
\begin{eqnarray}
\label{eq:diagonalization}
U^T\,\mathcal{M}\,U = \mathrm{diag}\{m_h^2-\delta m_h^2,\ m_h^2+\delta
m_h^2,\ m_{h_3}^2\}. 
\end{eqnarray}
To simplify the expressions obtained from eq.~(\ref{eq:diagonalization}) we
factorise $c_{12}$ and $c_{23}$ to write U in terms of
$t_{kl}\equiv\tan\theta_{kl}$ and use the approximations:
\begin{eqnarray}
\label{eq:Uapprox2}
c_{12}\simeq 1,\qquad c_{23}\simeq 1, \qquad \frac{1}{2}\pm\tan\theta_{kl}\tan\theta_{mn}\simeq \frac{1}{2},
\end{eqnarray}
where $kl=12,23$ and $mn=12,23$. Finally, we will focus on 
the relations 
in terms of the mass matrix elements $\mathcal{M}_{22}$
  and $\mathcal{M}_{23}$ since $\mathcal{M}_{22}^{\mathrm{tree}}$ and
  $\mathcal{M}_{23}^{\mathrm{tree}}$ reproduce pretty well the values computed
  by NMSSMtools, and because we wish to get simple relations between the Higgs
  sector parameters, masses and mixing angles. %
  We have checked numerically that for the rest of mass matrix elements the
  tree level expression are not precise enough.
\begin{eqnarray}
  \label{eq:mapprox02}
  \mathcal{M}_{22} - m_{h_3}^2 &=&
               m_h^2\left(t_{23}^2+t_{12}^2\right) +
               \cos(2\theta_{13})\,\delta m_h^2
              \left(t_{23}^2-t_{12}^2\right) %
                                   + 2\sin(2\theta_{13})\,\delta m_h^2 t_{12}\,t_{23} \\ %
  \label{eq:mapprox00}
   \mathcal{M}_{13} + \mathcal{M}_{23}\, t_{12} &=& -\sin(2\theta_{13})\,\delta
                                                    m_h^2 %
  \\ \label{eq:mapprox01}
  \mathcal{M}_{22}\,t_{23} - \mathcal{M}_{23} &=&
  m_h^2 t_{23} + \delta m_h^2 \left[\, t_{12}\sin(2\theta_{13}) + t_{23}\cos(2\theta_{13})\, \right].
\end{eqnarray}
We can further simplify eq.~(\ref{eq:mapprox02}) taking into account that
$\delta m_h^2$ and $m_h^2$ are smaller than $m_{h_3}$ and $\mathcal{M}_{22}$. 
Using the last approximation of eq.~(\ref{eq:Uapprox2}) we get that terms
proportional to $t_{12}^2$, $t_{23}^2$ and $t_{12}t_{23}$ in the right hand of
eq.~(\ref{eq:mapprox02}) are negligible. Regarding eq.~(\ref{eq:mapprox01}),
using the approximation eq.~(\ref{eq:Uapprox2}) and eq.~(\ref{eq:mapprox00})
one gets
$\mathcal{M}_{23}+\delta m_h^2\,t_{12}\sin(2\theta_{13})\simeq
\mathcal{M}_{23}$, allowing us to neglect the term proportional to $t_{12}$ in
eq.~(\ref{eq:mapprox01}) (besides that, for the sample of
pNMSSM 
points described in section~\ref{sec:scan} the values of $\theta_{12}$ are
much smaller than the values of $\theta_{23}$). %
Hence eqs.~(\ref{eq:mapprox02})-(\ref{eq:mapprox01}) can be rearranged to
get,
\begin{eqnarray}
  \label{eq:mapprox1}
  m_{h_3}^2 &=& \mathcal{M}_{22},\\   \label{eq:mapprox2}
  t_{12} &=& -\frac{\sin(2\theta_{13})\,\delta m_h^2 +
                  \mathcal{M}_{13}}{\mathcal{M}_{23}}, \\   \label{eq:mapprox3}
  t_{23} &=& \frac{\mathcal{M}_{23}}{m_{h_3}^2-m_h^2 - %
             \delta m_h^2 \cos(2\theta_{13})} %
             \simeq  \frac{\mathcal{M}_{23}}{m_{h_3}^2-m_h^2}, 
\end{eqnarray}
where in the last equation we have further considered that
$\delta m_h^2 \cos(2\theta) \ll m_h^2$.\\

\noindent Using the approximation of large $\tan\beta$ and large
$M_A$ from reference~\cite{Miller:2003ay}:
\begin{displaymath}
  m_{h_3}^2\approx M_A^2(1+\frac{1}{4}\frac{\lambda v}{\mu}\sin^2{2\beta}).
\end{displaymath}
We have checked numerically that $m_{h_3}\approx M_A$ is a good approximation
for the pNMSSM points considered.  
Now, let us take $\mathcal{M}_{23}$ from reference \cite{Miller:2003ay}%
  \footnote{Since they perform a different rotation, written
  in eq.~16 of \cite{Miller:2003ay}, we transform the mass matrix as follow:%
 \begin{eqnarray}\nonumber
   \label{eq:massmattrans}
   \mathcal{M}= \left(\begin{array}[h]{ccc}
                        M_{22} & -M_{21} & M_{23}\\
                        -M_{12} & M_{11} & -M_{13}\\
                        M_{32} & -M_{31} & M_{33}
   \end{array}\right)
 \end{eqnarray}} %
\begin{displaymath}
  \mathcal{M}_{23}=\frac{1}{2}\frac{v}{v_s}\cos{2\beta}\left(M_A^2\sin{2\beta}+\lambda\kappa v_s^2\right)
\end{displaymath}
and replace it in eq.~(\ref{eq:mapprox3}), considering
that $M_A$ is much heavy than $m_h^2$ one can write $t_{23}$ as,
\begin{eqnarray}
  \label{eq:t23approx}
  t_{23}\simeq \frac{\lambda}{2} \cos{2\beta}\, \sin{2\beta}
  \left(\frac{v}{\sqrt{2}\, \mu} \right)
  \left(1+2\,\frac{\kappa}{\lambda}\,\frac{\xi^2}{\sin{2\beta}} \right)
\end{eqnarray}
where $v_s=\sqrt{2}\,\mu/\lambda$ and $\xi=\mu/M_A$.%
\begin{figure}[t!]
  \centering
  \includegraphics[width=0.45\textwidth]{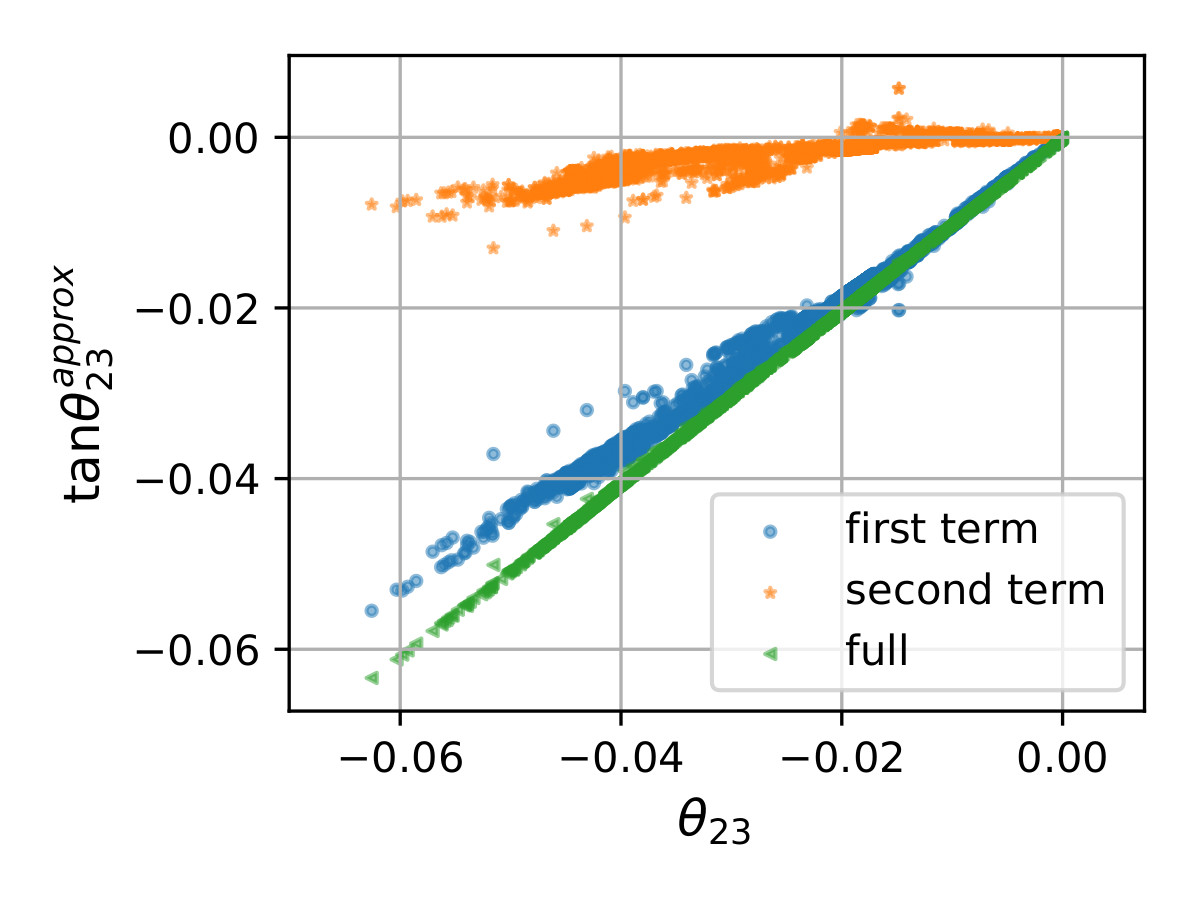}  \,\,
  \includegraphics[width=0.45\textwidth]{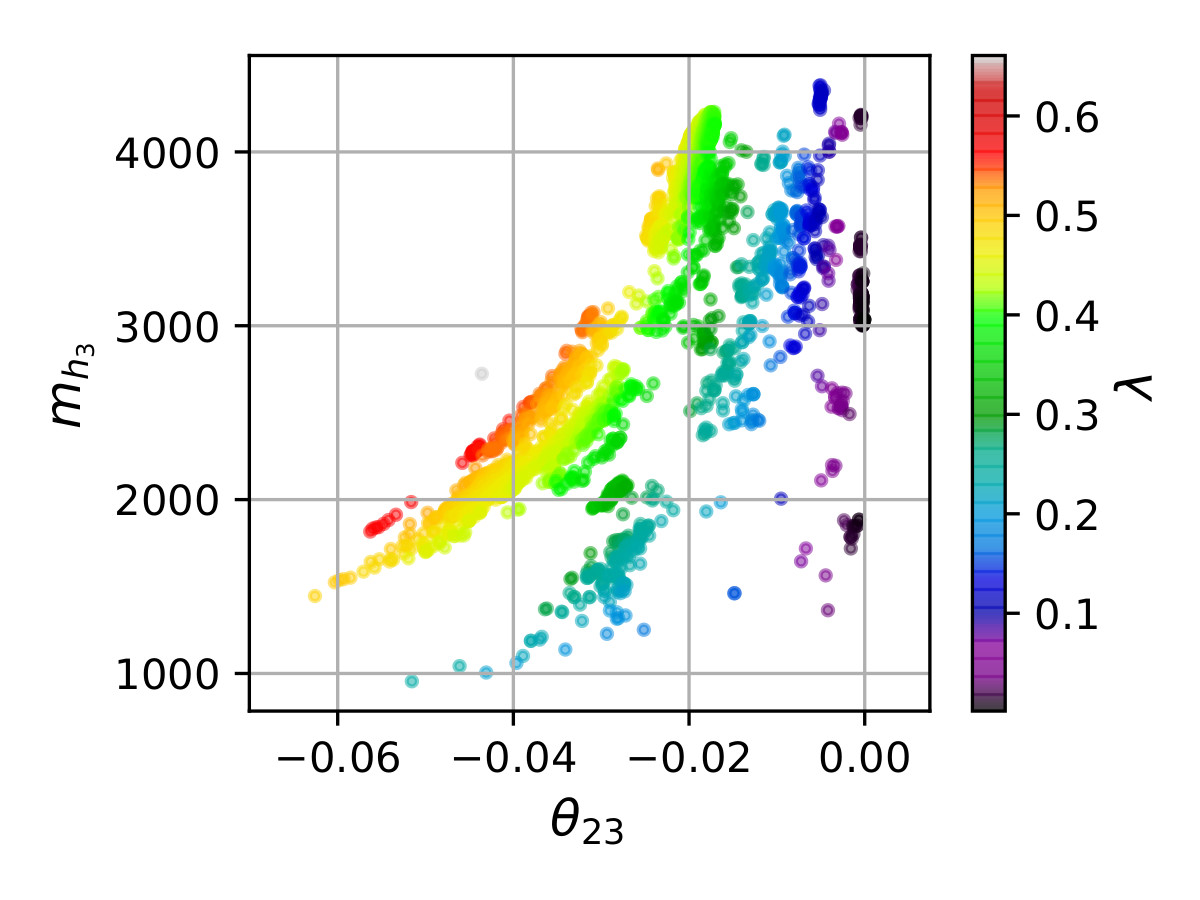}  \,\,
  \caption{Left panel: $\tan\theta_{23}$ approximation showed in equation
    (\ref{eq:t23approx}). Right panel: The mass of the heaviest CP-even
      Higgs as function of $\theta_{13}$ and $\lambda$.}
  \label{fig:t23approx}
\end{figure}

Left panel of Figure~\ref{fig:t23approx} shows in the x-axes the value of
$\theta_{23}$ computed by NMSSMtools and in the y-axes the analytical
approximation described in eq.~(\ref{eq:t23approx}), as one can see in the
Figure there is a good agreement between the analytical expression and the
numerical value (green points), and it is clear that the main contribution to
$\theta_{23}$ comes from the first term of eq.~(\ref{eq:t23approx}) (blue
points). %
Right panel of Figure~\ref{fig:t23approx} shows the relation between
$\theta_{23}$ and $m_{h_3}$ for constant values of $\lambda$. There is a
trend: larger values of $|\theta_{23}|$ correspond to smaller values of
$m_{h_3}$, except for very small values of $|\theta_{23}|$ where the two
parameters seem to be uncorrelated. Still, eq.~(\ref{eq:t23approx})
shows that the value of $\tan{\theta_{23}}$ is not directly related to the
scale of the heaviest Higgs, but instead it is related to the value of
$\lambda$, $\mu$ and $\tan\beta$ 
\footnote{Let us remember that in the decoupling limit of $H_0$,
\begin{displaymath}
  m_{h_3}^2 \simeq M_A^2 = \frac{2 \mu}{\sin{2\beta}}\left(A_{\lambda}+\frac{\kappa}{\lambda}\mu\right)
\end{displaymath}
}. \\

Although the Higgs boson masses get 
important contributions from loop corrections, it is possible to get some
information from the tree level expressions for $m_{h_{1}}$ and
$m_{h_{2}}$. For 
large values of $\tan\beta$ and $M_A^2$, 
\begin{eqnarray}\nonumber
  \label{eq:mh1mh2}
  [m_{h_{2/1}}^2]^{\mathrm{tree}}&=&\frac{1}{2}\left\{ M_Z^2 +
    \frac{1}{2}\kappa v_s(4\kappa v_s + \sqrt{2}A_\kappa) \right. \\
&& \qquad \left. \pm \sqrt{\left[ M_Z^2-\frac{1}{2}\kappa v_s(4\kappa
  v_s+\sqrt{2} A_\kappa) \right]^2 + \frac{v^2}{v_s^2}\left[2\lambda^2v_s^2 - M_A^2{\sin{2\beta}}^2\right]^2} \right\}
\end{eqnarray}
where $v_S=\sqrt{2}\mu/\lambda$ (see Eq.~(32) of \cite{Miller:2003ay}). %
In order to get a constrain for the initial parameters from the
  condition of small mass difference between the two lightest Higgs states, we
  require a small mass difference between the tree level masses showed in
  eq.~(\ref{eq:mh1mh2}). %
  But, since the tree level expression do not precisely reproduce the masses
  of the Higgs states we request the mass square difference at tree level to
  be smaller than $M_Z^2$, meaning that both terms
  inside the square root should be smaller than $M_Z^4$.\\
Let us focus on the first term, for $A_\kappa\gg M_Z$ there should be a
correlation between $A_\kappa$ and $\kappa v_s$ such that there is a
cancellation that leads to an order $M_Z^2$ value. Note that the average of
the tree-level squared masses also requires this cancellation to occur in
order to get the masses of the Higgs states in the desired range.\\
For $|A_\kappa| \gg M_Z$ we expect,
\begin{eqnarray}
  \label{eq:AkappaRelation}
  A_\kappa\simeq -2\sqrt{2}\kappa v_s.
\end{eqnarray}
\begin{figure}[t!]
  \centering
  \includegraphics[width=0.45\textwidth]{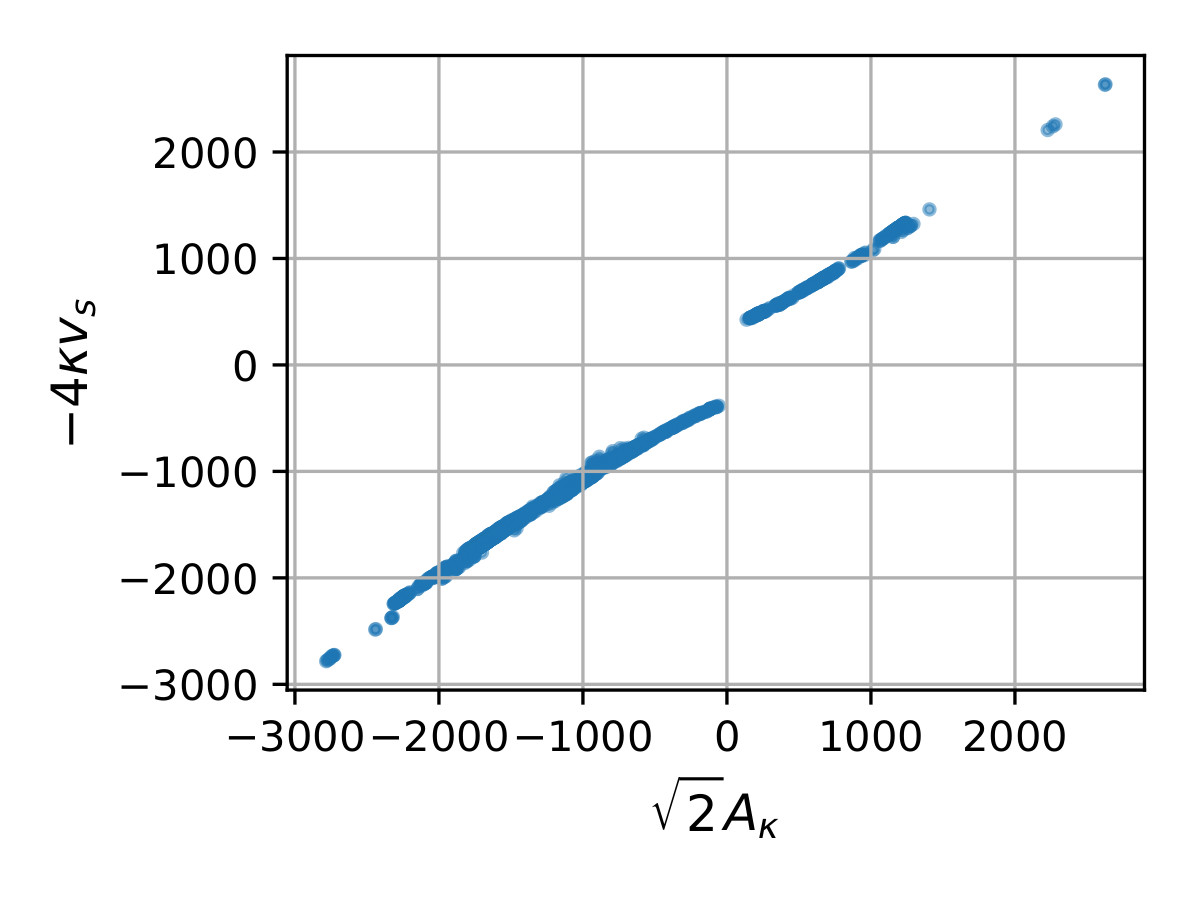} \,\, %
    \includegraphics[width=0.45\textwidth]{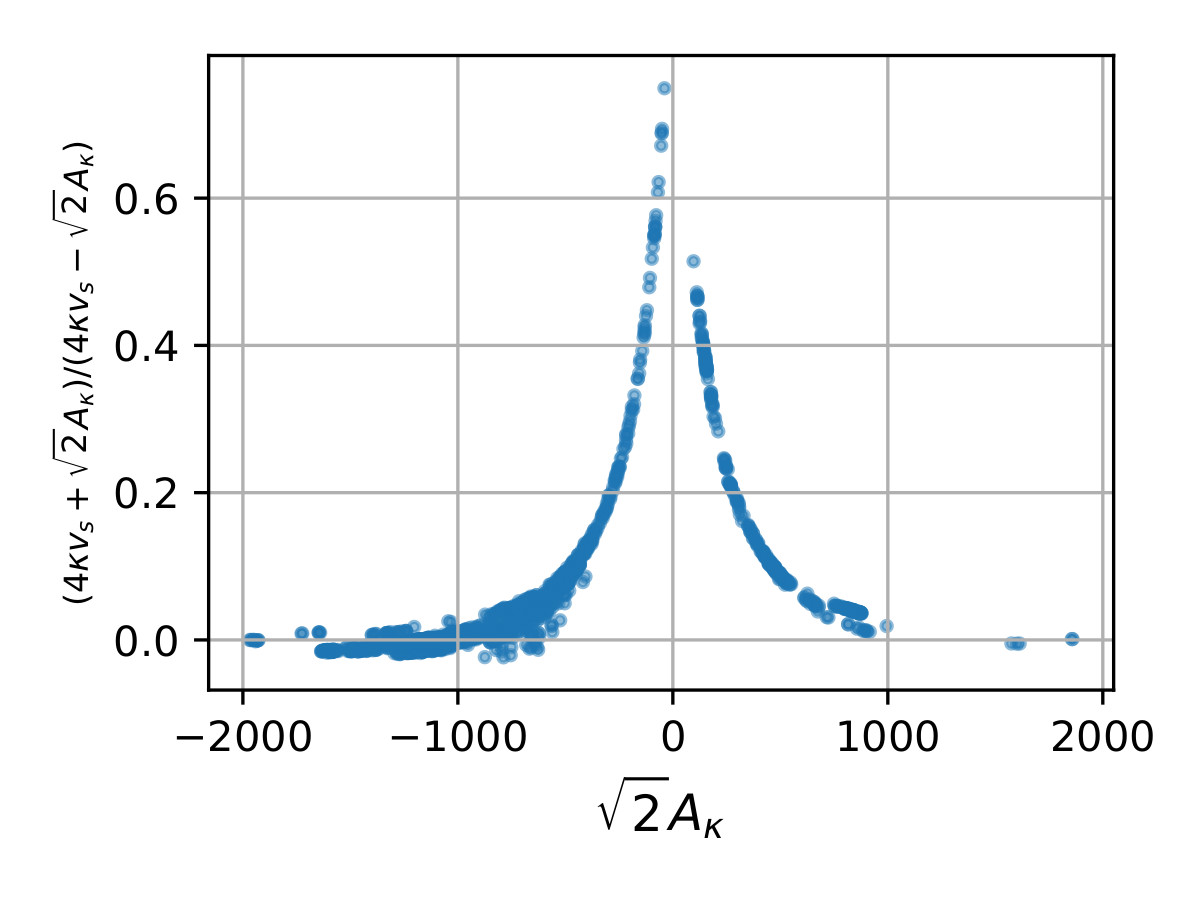}
    \caption{Left panel shows the relation between $A_\kappa$ and
        $\kappa v_s$.  Right panel shows the relation between $A_\kappa$ and
        the degree of cancellation of $4\kappa v_s+\sqrt{2} A_\kappa$ for the 
        sample of pNMSSM 
        points considered.}
  \label{fig:Akappavs}
\end{figure}
Figure~\ref{fig:Akappavs} shows the relation between $A_\kappa$ and $\kappa
v_s$, as manifested in the figure for $|A_\kappa|\gtrsim 600$~GeV the
approximation of eq.~(\ref{eq:AkappaRelation}) works within an error
smaller than $5\%$.\\

Furthermore, using eq.~(\ref{eq:AkappaRelation}) it is possible to simplify
other parameters relevant in the Higgs sector, eq.~(30) of
\cite{Miller:2003ay} gives a simplified expression for the mass of the light
pseudoscalar,
\begin{eqnarray}
  \label{eq:mA1}
  m_{A_1}^2 &\simeq& -\frac{3}{\sqrt{2}}\kappa v_s A_\kappa.
\end{eqnarray}
Putting eq.~(\ref{eq:AkappaRelation}) into eq.~(\ref{eq:mA1}) we write the
mass of the lightest pseudoscalar in terms of $\kappa$ and $v_s$,
\begin{eqnarray}
  \label{eq:mA1approx}
  m_{A_1}^2 &\simeq& 6\, \kappa^2\, v_s^2.
\end{eqnarray}
\begin{figure}[t!]
  \centering
  \includegraphics[width=0.5\textwidth]{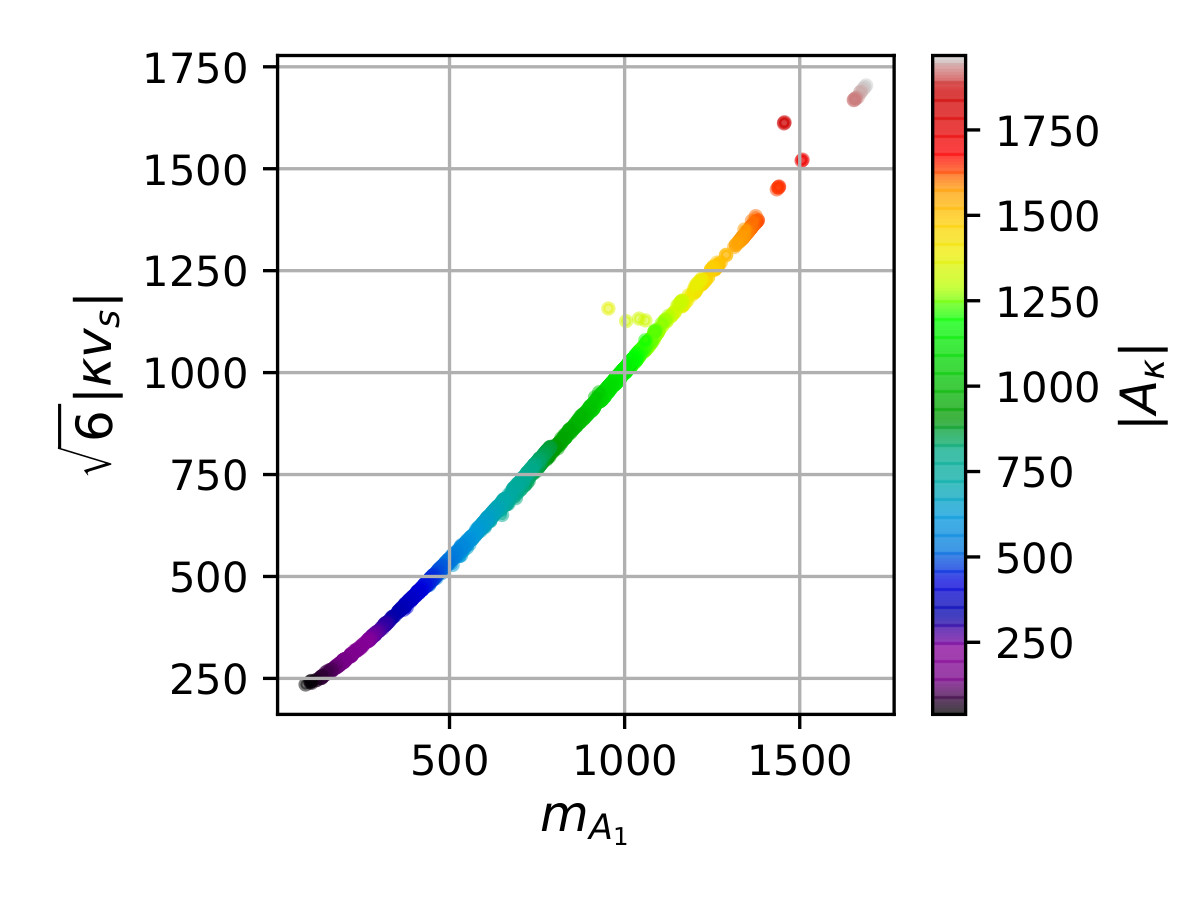}
  \caption{Comparison between the mass of the lightest pseudoscalar
      computed by NMSSMtools and the approximate analytical value described in
      eq.~(\ref{eq:mA1approx}). The colour code shows the value of $A_\kappa$,
      which as described in eq.~(\ref{eq:AkappaRelation}) it is related with
      the value of $\kappa v_s$.}
  \label{fig:mA1approx}
\end{figure}
Figure~\ref{fig:mA1approx} shows the comparison between
eq.~(\ref{eq:mA1approx}) and the value computed by NMSSMtools. It can be seen that 
for $m_{A_1}>500$~GeV eq.~(\ref{eq:mA1approx}) is a pretty
good representation for the light pseudoscalar mass. \\

For completeness, it is worth mentioning that 
the second term inside the squared root of eq.~(\ref{eq:mh1mh2}) is
suppressed by a factor $v_s^{\,-2}$, as such we do not 
expect to get any good correlation of parameters from there.\\

\noindent All the information, presented above, are useful for determining an
optimal range of parameters in order to perform a specialised parameters scan dedicated
for studying mass-degenerate Higgs region(s).\\

\section{The two lightest CP-even Higgses at the LHC}
\label{sec:2cpevenh}

In this section we will use the results of the scan and the analytical
relations for the couplings and signal strengths to study the parameter space
where the two lightest CP-even Higgs states mimic the SM-Higgs signals.\\

First, we have to verify the validity of the analytic expressions for the
signal strengths comparing these expressions with the numerical values
computed by NMSSMtools.\footnote{To perform this comparison we flip the order
  of the mass eigenstates computed by NMSSMtools, in such a way that $h_1$ has
  the largest component of $h_0$, and it is not necessary the lightest mass
  eigenstate. The need of this transformation is due to the convention used
  for the Higgs mixing matrix in NMSSMtools. The determinant of this matrix
  could be positive or negative depending on $h_0$-fraction of the lightest
  eigenstate. It is positive if $h_1$ is $h_0$-dominated and negative if it is
  $S$-dominated.\\ The reason why we perform the flip of states is because we
  want to make a comparison of the analytic relations as function of the
  mixing angles, for this we need to assume a specific form of the mixing
  matrix U.} %
\begin{figure}[t!]
  \centering
  \includegraphics[width=0.45\textwidth]{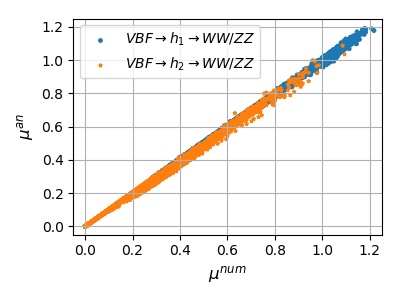} %
  \includegraphics[width=0.45\textwidth]{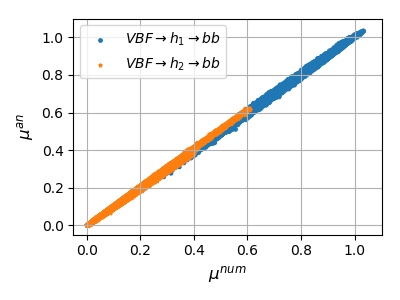}
  \caption{Shows the $\mu$ factor for VBF$\rightarrow h_i \rightarrow WW/ZZ$
    (left panel), and VBF$\rightarrow h_i \rightarrow bb$ (right panel) for
    the pNMSSM posterior sample considered which has 
    $m_{h_2}-m_{h_1}<3$~GeV with both $m_{h_1}$ and $m_{h_2}$ allowed within [$122$, $128$] GeV.}
  \label{fig:mucomparison}
\end{figure}
Figure~\ref{fig:mucomparison} shows the comparison between the signal
strengths computed by NMSSMtools, $\mu^{\mathrm{num}}$, and the analytic
approximations showed in eqs.~(\ref{eq:mutot1})-(\ref{eq:mutot4}),
$\mu^{\mathrm{an}}$, for VBF$\rightarrow h_i \rightarrow WW/ZZ$ (left
panel) and VBF$\rightarrow h_i \rightarrow bb$ (right panel). %
From the figure we see that there is a good agreement between the analytical approximation and the numerical computation.\\

Now, let us identify the relevant parameters that produce deviation from
experimental measurement. %
Writing the couplings, widths and signal strengths in
terms of the mixing angles, for small values of $\theta_{12}$ and
$\theta_{23}$, see eqs.~(\ref{eq:couplings}) and (\ref{eq:Umixmat2haprox}),
\begin{eqnarray}
  \label{eq:hcouplings2}
  && g_{h_ib_Lb_R^c} \simeq %
  \frac{m_b}{\sqrt{2}v} \left\{ \begin{array}[h]{cc}
                                  c_{13}+(c_{13}\theta_{12}-s_{13}\theta_{23})\tan\beta, & i=1 \\
                                  -s_{13}-(s_{13}\theta_{12}+c_{13}\theta_{23})\tan\beta, & i=2 
                                \end{array} \right.\\ \nonumber
  \\
  && g_{h_iZ_\mu Z_\nu} \simeq %
   g_{\mu\nu}\frac{g_1^2+g_2^2}{\sqrt{2}}v %
  \left\{ \begin{array}[h]{cc}
            c_{13}, & i=1\\
            -s_{13}, & i=2 \end{array} \right.  \\ \nonumber
  \\
  && g_{h_iW_\mu^+ W_\nu^-} \simeq%
   g_{\mu\nu}\frac{g_2^2}{\sqrt{2}}v %
  \left\{ \begin{array}[h]{cc}
            c_{13}, & i=1\\
            -s_{13}, & i=2 \end{array} \right. .
\end{eqnarray}
Using eq.~(\ref{eq:hcouplings2}) and eq.~(\ref{eq:width}) we get,
\begin{eqnarray}
  \label{eq:widthh1h2a}
  \Gamma_1/\Gamma_{SM} &=& (1-\mathrm{BR}_{bb})(c_{13})^2 +
                                   \mathrm{BR}_{bb}[c_{13} +
                                   c_{13}\theta_{12}\tan\beta -
                                   s_{13}\theta_{23}\tan\beta]^2 \\ \textrm{ and }
                                   \label{eq:widthh1h2b}
  \Gamma_2/\Gamma_{SM}&=& (1-\mathrm{BR}_{bb})(s_{13})^2 +
                                   \mathrm{BR}_{bb}
                                   [s_{13} + s_{13}\theta_{12}\tan\beta + c_{13}\theta_{23}\tan\beta]^2.
\end{eqnarray}
Finally, eq.~(\ref{eq:mutot1})-(\ref{eq:mutot2}) can be written in terms of
the mixing angles as 
\begin{eqnarray}
  \label{eq:muan1}
  \mu_{VBF\rightarrow h_1 \rightarrow WW/ZZ}^{\mathrm{an}} &\simeq&
  \frac{(c_{13})^4}{(1-\mathrm{BR}_{bb})(c_{13})^2 + 
                            \mathrm{BR}_{bb}(c_{13} +
                            c_{13}\theta_{12}\tan\beta -
                                     s_{13}\theta_{23}\tan\beta)^2}, \\   \label{eq:muan2}
  \mu_{VBF\rightarrow h_2 \rightarrow WW/ZZ}^{\mathrm{an}} &\simeq&
  \frac{(s_{13})^4}{(1-\mathrm{BR}_{bb})(s_{13})^2 + 
                            \mathrm{BR}_{bb}
                            (s_{13} + s_{13}\theta_{12}\tan\beta +
                                     c_{13}\theta_{23}\tan\beta)^2}, \\ \label{eq:muan3}
  \mu_{VBF\rightarrow h_1 \rightarrow bb}^{\mathrm{an}} &\simeq& \frac{(c_{13})^2(c_{13}+c_{13}\theta_{12}\tan\beta+s_{13}\theta_{23}\tan\beta)^2}{(1-\mathrm{BR}_{bb})(c_{13})^2 +
                            \mathrm{BR}_{bb}(c_{13} +
                            c_{13}\theta_{12}\tan\beta -
                                     s_{13}\theta_{23}\tan\beta)^2}, \textrm{ and }\\ \label{eq:muan4}
  \mu_{VBF\rightarrow h_2 \rightarrow bb}^{\mathrm{an}} &\simeq& \frac{(s_{13})^2  (s_{13}+s_{13}\theta_{12}\tan\beta+c_{13}\theta_{23}\tan\beta)^2}{(1-\mathrm{BR}_{bb})(s_{13})^2 + 
                            \mathrm{BR}_{bb}
                            (s_{13} + s_{13}\theta_{12}\tan\beta + c_{13}\theta_{23}\tan\beta)^2}.
\end{eqnarray}
From eqs.(\ref{eq:muan1})-(\ref{eq:muan4}) we see that the signal strengths
depend on four parameters: $\theta_{13}$, $\theta_{23}$, $\theta_{12}$ and
$\tan\beta$. However, in the limit where $\theta_{12}\tan\beta\ll\theta_{13}$,
which is the case for the pNMSSM posterior sample analysed, 
the number of parameters reduces to two:%
\begin{displaymath}
  \theta_{13}\ , \qquad \theta_{23}\tan\beta.
\end{displaymath}
From eqs.~(\ref{eq:muan1})-(\ref{eq:muan4}), one can see that the dependence on $\theta_{12}$ always appears as a factor
in the expression $\cos{\theta_{13}}(1+\theta_{12} \tan\beta)$ or $\sin{\theta_{13}}(1+\theta_{12} \tan\beta)$. Therefore for
$\theta_{12}\tan\beta \ll 1$ 
the contribution of $\theta_{12}$ is negligible.\\

\begin{figure}[t!]
  \centering
  \includegraphics[width=0.45\textwidth]{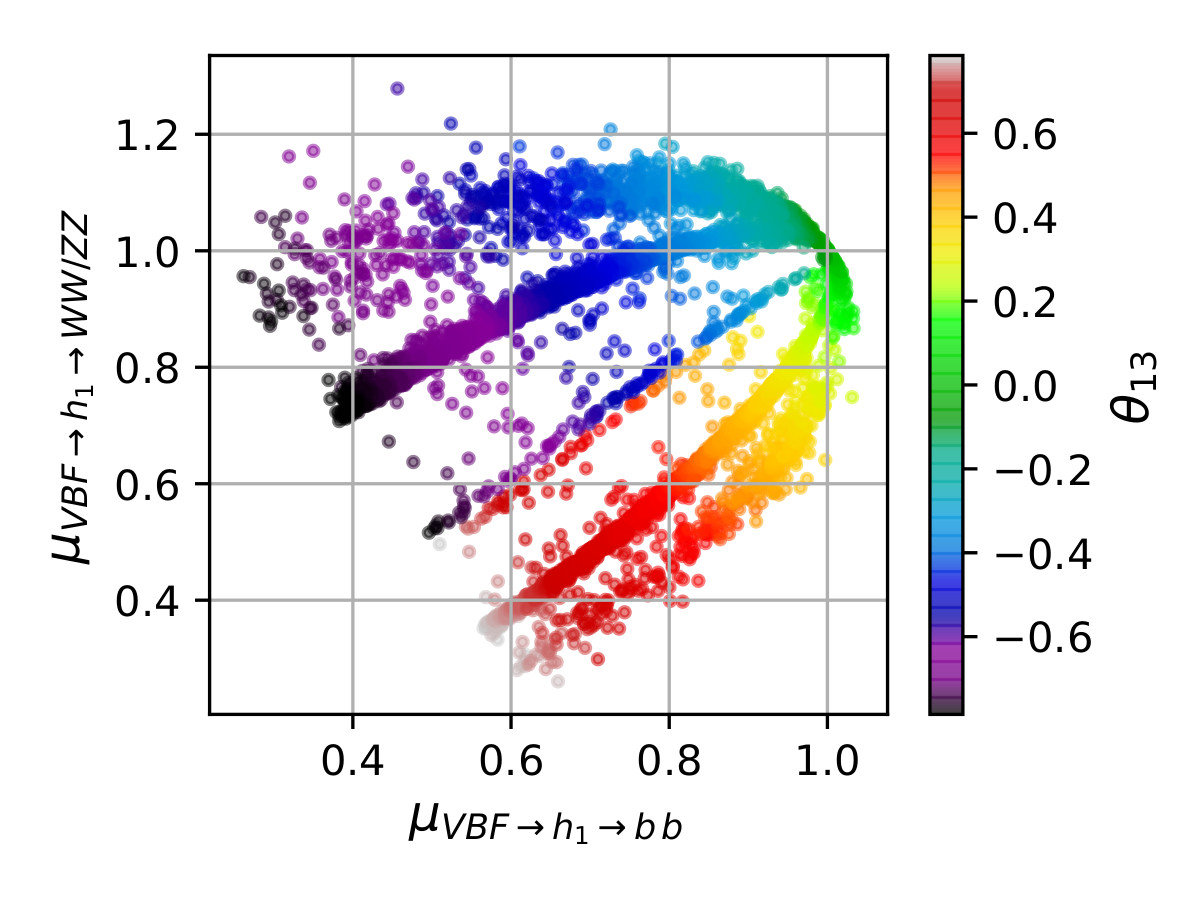}%
  \includegraphics[width=0.45\textwidth]{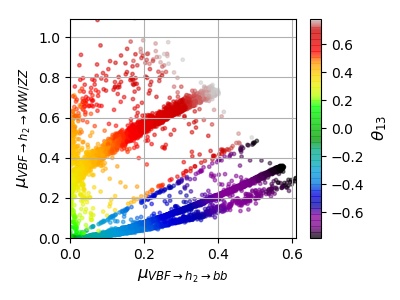}\\
  \includegraphics[width=0.45\textwidth]{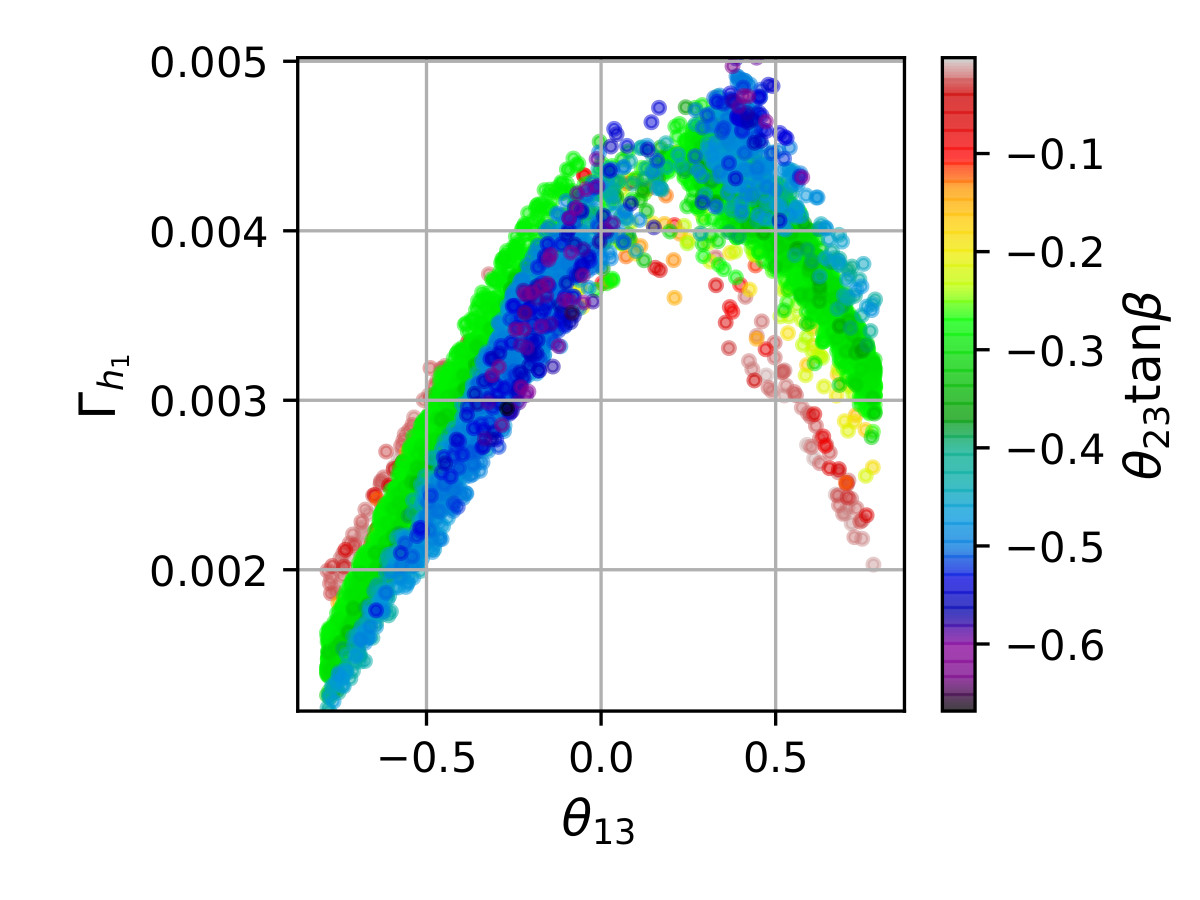}%
  \includegraphics[width=0.45\textwidth]{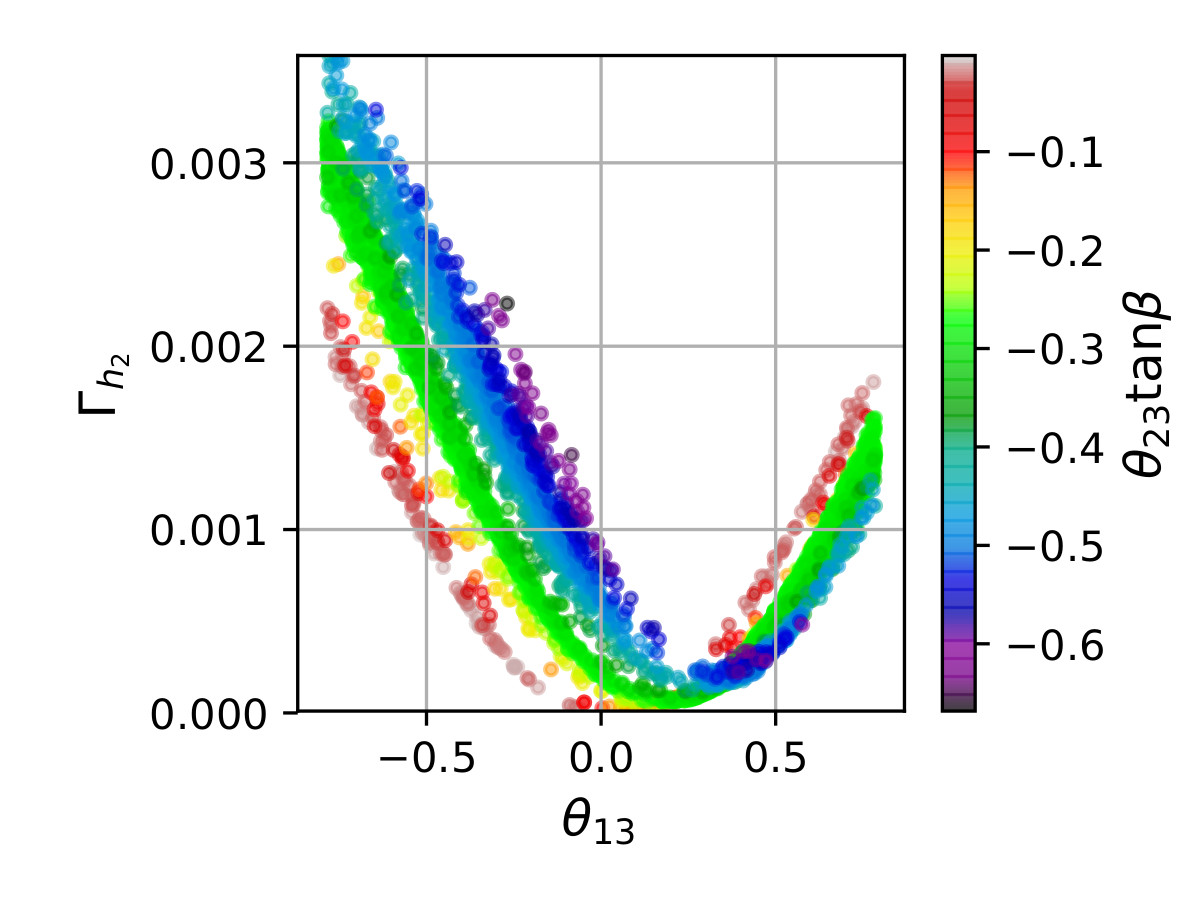}
  \caption{Top panels show the relation between both $\mu$ factors in
      terms of $\theta_{13}$ for $h_1$ (top left) and $h_2$(top right), the
      colour code shows the dependence on $\theta_{13}$. Bottom
      row of plots show the width of $h_1$ (bottom left) and $h_2$ (bottom
      right) as function of $\theta_{13}$. The colour code shows the dependence with 
      respect to $\theta_{23}\tan\beta$.}
  \label{fig:muWWZZbb}
\end{figure}
To understand the dependence of the signal strengths with respect to
$\theta_{13}$ and $\theta_{23}\tan\beta$ let us start analysing the relation
between the signal strengths for a given Higgs state. The top row of
Figure~\ref{fig:muWWZZbb} shows the correlations between
$\mu_{VBF\rightarrow h_i \rightarrow WW/ZZ}$ and
$\mu_{VBF\rightarrow h_i \rightarrow bb}$ for $h_1$ (top left) and $h_2$ (top
right); for $h_1$ we can see that the difference between the $b\bar{b}$ and
$WW/ZZ$ channel signal strengths is not small. In fact, this could be taken to
imply 
that it is not possible to reproduce the experimental results 
with such differences. However, looking at the right panel of the Figure
  and using the colour code to select regions with constant values of
  $\theta_{13}$, it is possible to compare the rates of the signal strengths
  for both Higgs bosons. The plots show that the enhancement(suppression) of one
channels of $h_1$ is more or less compensated with a
suppression(enhancement) in the same channel of $h_2$.\\

The  analytic
expressions for the widths of the Higgs states, eqs. (\ref{eq:widthh1h2a}) and
(\ref{eq:widthh1h2b}), show that the term proportional to
$\theta_{23}\tan\beta$ has a minus sign in the width of $h_1$ and plus sign in
the width of $h_2$, decreasing(increasing) the decay rate of
$h_1\rightarrow bb$ while increasing(decreasing) the decay rate of
$h_2\rightarrow bb$ as $|\theta_{23}|$ increases its value.\\
The bottom row of Figure \ref{fig:muWWZZbb} shows the width of $h_1$ and $h_2$
as function of $\theta_{13}$ and $\theta_{23}\tan\beta$. 
The figure agrees with what we expected from the approximate expressions,
eqs.(\ref{eq:widthh1h2a}) and (\ref{eq:widthh1h2b}), a function dominated by
$\cos^2{\theta_{13}}$ for $h_1$ and $\sin^2{\theta_{13}}$ for
$h_2$, the phase of the distributions varies with the values of
$|\theta_{23}\tan\beta|$.\\ 

\begin{figure}[t!]
  \centering
  \includegraphics[width=0.45\textwidth]{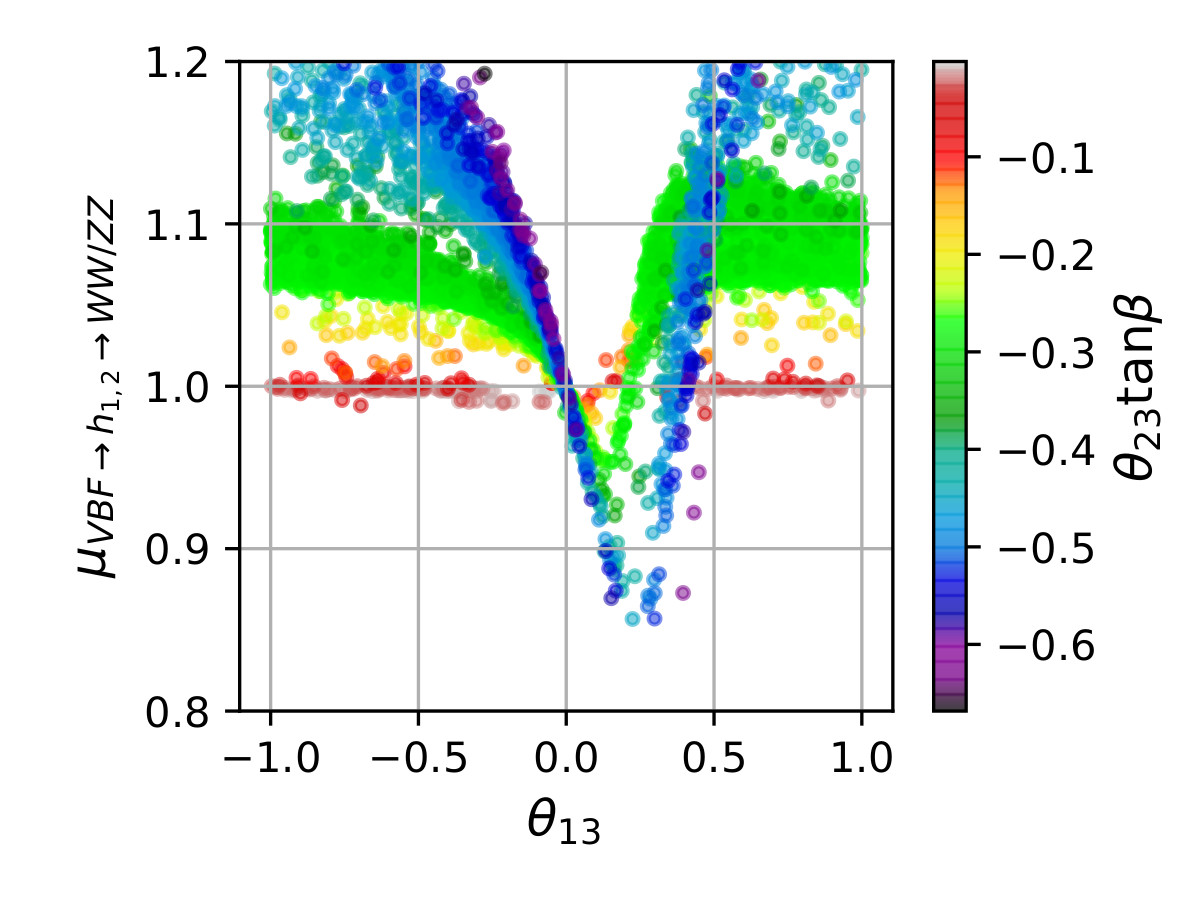}%
  \includegraphics[width=0.45\textwidth]{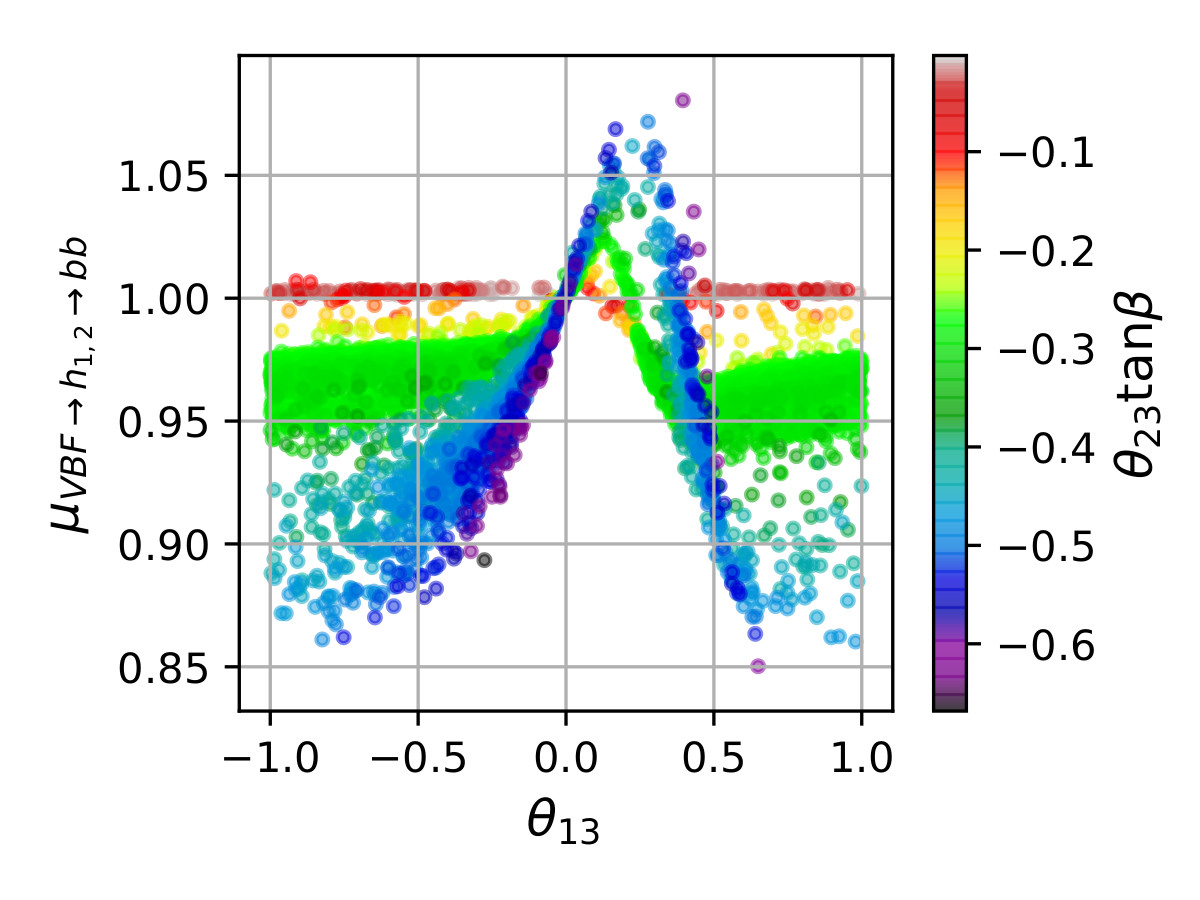}%
  \caption{Shows the signal strength of the superposition of the two
      Higgs states for vector-boson fusion production and WW/ZZ decay (left
      panel), and $bb$ decay (right panel). For both plots we show the
      dependence of the signal strength respect to $\theta_{13}$ and
      $\theta_{23}\tan\beta$.}
  \label{fig:mutotWWZZbb}
\end{figure}

Let us analyse the global signal strengths. Figure \ref{fig:mutotWWZZbb} shows
the sum of the signal strengths of vector-boson fusion production and decay to
$WW/ZZ$ (left panel) and to $bb$ (right panel), these factors represent the
global enhancement or suppression of the superposition of the two signals
respect to the signal of the standard Higgs. It is important to keep in mind
that to get the global signal strengths we sum the contributions of the 
individual signal strengths, which is allowed since we require the mass
difference of the two lightest CP-even Higgs states to be small enough not to
be resolved by current experiments, but much larger than the width of the
particles to neglect interference effects.\footnote{With this approach we are
  not considering the shape of the signal distribution. The analysis of the
  shape of the distribution goes beyond the scope of this work}\\

There are several points we would like to comment from Figure~\ref{fig:mutotWWZZbb}, the departure of the signal strength increases with the
size of $\theta_{23}\tan\beta$ as in the case of the individual signal
strengths. The modification of the signal strengths for $h_1$ is
``compensated'' by the modifications of the signal strengths for $h_2$ and
therefore the total effect is smaller than the one for the individual rates
but still not negligible. %
Regarding the relation between the two global signals strengths it is clear
from Figure \ref{fig:mutotWWZZbb} that $\mu_{VBF\rightarrow h_{1,2}
  \rightarrow WW/ZZ}$ has opposite behaviour and larger range with respect to
$\mu_{VBF\rightarrow h_{1,2} bb}$.\\ 

There are two regions that seem to be in full agreement with the SM (the
signal strength is $\simeq 1$): the region where $\theta_{23}\simeq 0$ and the
region where $\theta_{13}\simeq 0$, as we expected. There is a third region
where $\theta_{13}$ is between 0.2 and 0.4, where for a very precise value of
$\theta_{23}$ the signal strength is very close to one. On the other hand, for
small values of $\theta_{23}$, let's say $\theta_{23}\tan\beta\gtrsim -0.25$,
the deviation from one of the signal strength is very small, very precise
measurements will be necessary to resolve it.\\

There is one last comment about Figures \ref{fig:muWWZZbb}
and \ref{fig:mutotWWZZbb}. We are able to fully describe the rates and
the widths of $h_1$ and $h_2$ in terms of two parameters: $\theta_{13}$ and
$\theta_{23}\tan\beta$, instead of three, indicating that 
  $\theta_{12}\tan\beta\ll 1$ for the set of successful scanned points.
\\

So far we have focused our study to two channels:
VBF$\rightarrow h_i\rightarrow WW/ZZ$ and VBF$\rightarrow h_i\rightarrow bb$,
but the current measurements of the Higgs couplings constrain several more
channels. Let us comment about the most relevant production and decays:
\begin{enumerate}[a)]
\item Production processes like gluon-gluon fusion (GGF) and Higgs production
  associated to top quarks (ttH) are very important. To analyse these let us go 
  back to eqs.(\ref{eq:couplings}), which describe the couplings of the Higgs states to
  top quarks,
  \begin{displaymath}
    \hat{g}_{h_i t t}=U_{i1}-U_{i2}\cot\beta \simeq \left\{ %
      \begin{array}[h]{l r}%
        c_{13}-(c_{13}\theta_{12}-s_{13}\theta_{23})\cot\beta, & i=1\\
        -s_{13}+(s_{13}\theta_{12}+c_{13}\theta_{23})\cot\beta, & i=2%
      \end{array} \right..
  \end{displaymath}
  Comparing $\hat{g}_{g_i tt}$ with $\hat{g}_{g_i bb}$ we see that the
  contribution from $\theta_{23}$ is $\cot^2\beta$ times smaller for
  $\hat{g}_{h_i tt}$ than for $\hat{g}_{h_i bb}$, therefore we expect the
  contribution of $\theta_{23}$ to be very tiny and the production processes
  of GGF and ttH to behave as vector-boson fusion for given values of
  $\theta_{13}$ and $\theta_{23}\tan\beta$. \\

\item The Higgs decay to photons was one of the most important channels for
  the discovery of a new particle, where the main contribution to the decay of
  the standard Higgs to photons is through a loop of W bosons. We expect that
  the decay of the Higgs states to photons with respect to the value of the standard
  Higgs scale as the decay to WW/ZZ.

\item The decay of the Higgs states to taus with respect to the value of the
  standard Higgs will scale as the decay of the Higgs states to bottom quarks.

\end{enumerate}

\begin{figure}[t!]
  \centering
  \includegraphics[width=0.45\textwidth]{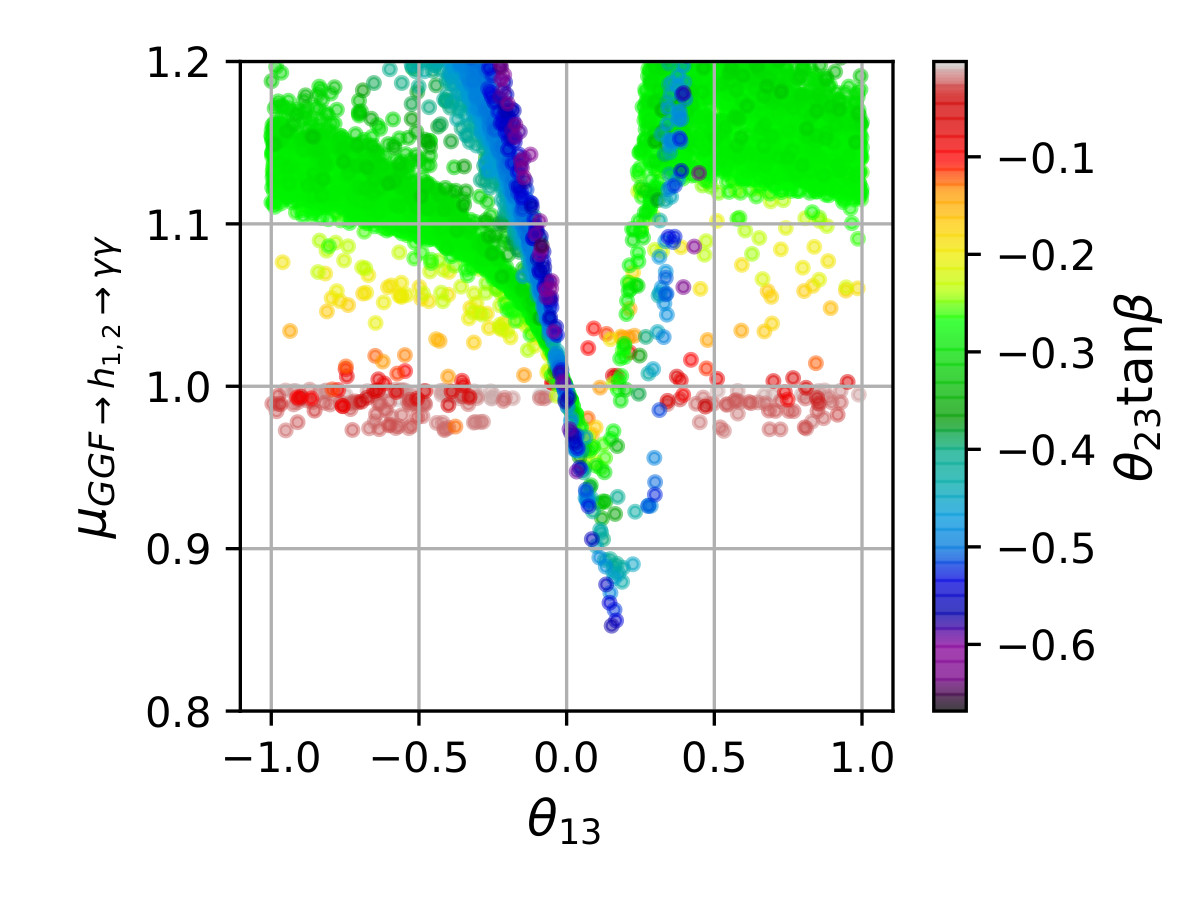}%
  \includegraphics[width=0.45\textwidth]{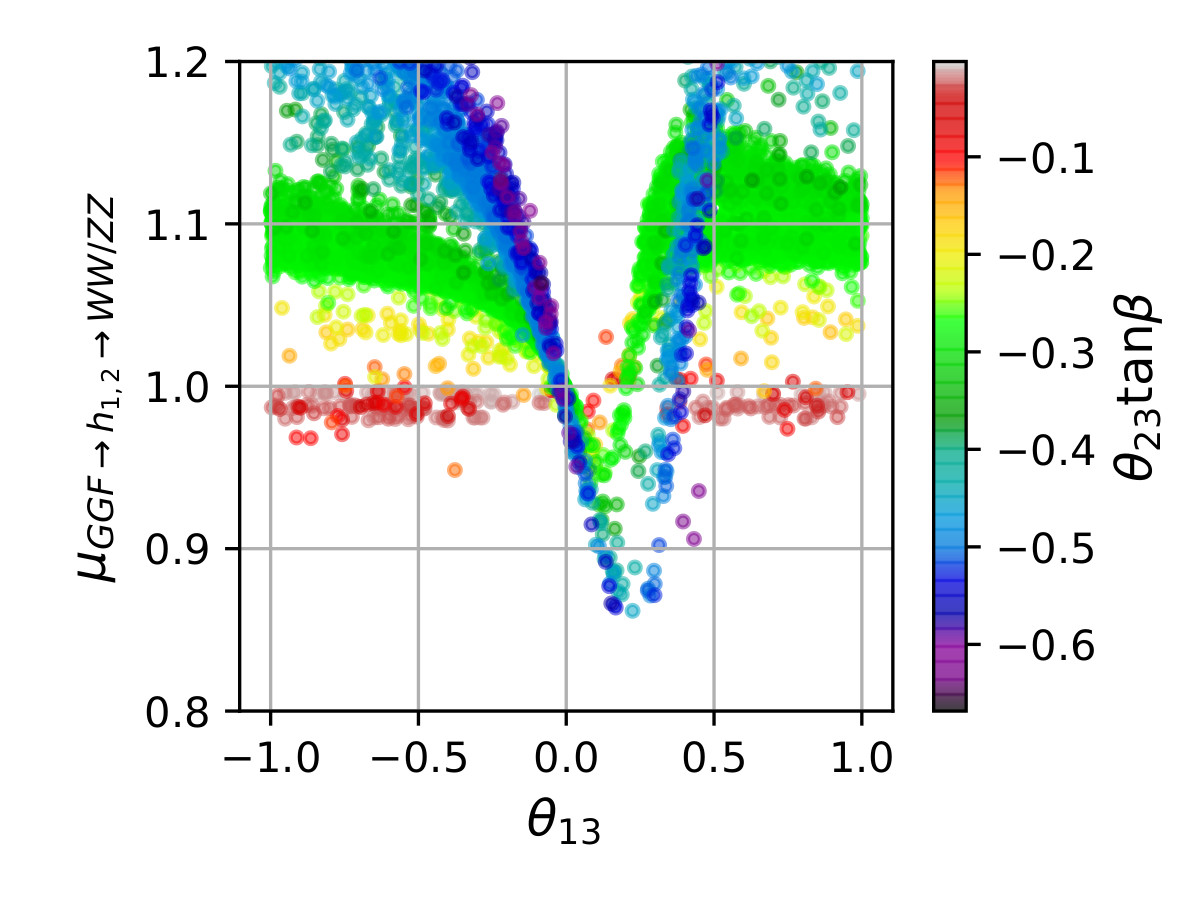}%
  \caption{Signal strengths for gluon-gluon fusion production processes
      and $\gamma\gamma$ decay (left panel), and $WW/ZZ$ decay (right panel).}
  \label{fig:muggf}
\end{figure}

To complete the description of the signals of the two lightest CP-even Higgs
states, in Figure~\ref{fig:muggf} we show the signal strengths for
GGF$\rightarrow h_{1,2}\rightarrow WW/ZZ$ (left panel) and
GGF$\rightarrow h_{1,2}\rightarrow \gamma\gamma$ (right panel). As we
expected, the gluon-gluon fusion production of the Higgses and decay to WW/ZZ
is pretty similar to the vector-boson fusion production, on the other hand,
the decay to photons shows a larger departure. 
%
\vspace{0.5cm}\\

\begin{figure}[t!]
  \centering
  \includegraphics[width=0.45\textwidth]{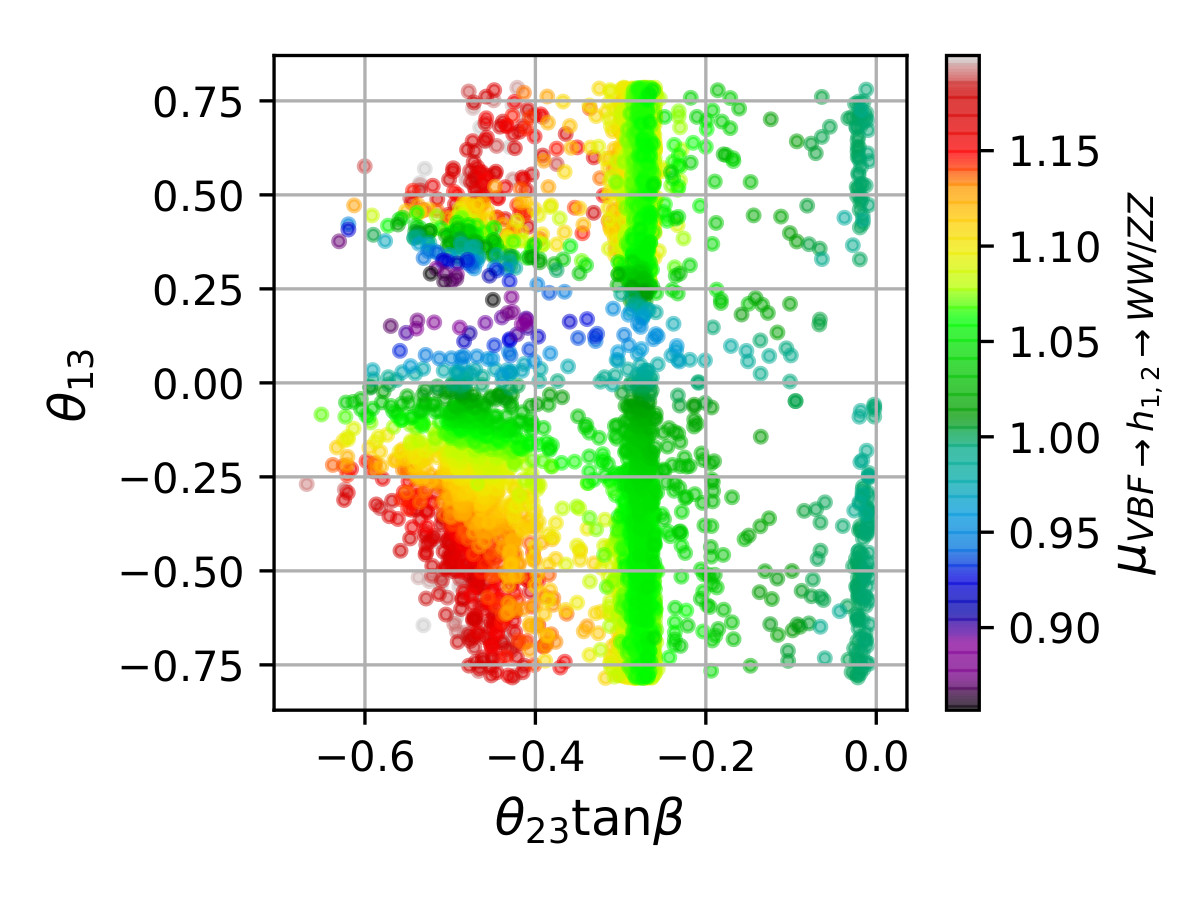} %
  \includegraphics[width=0.45\textwidth]{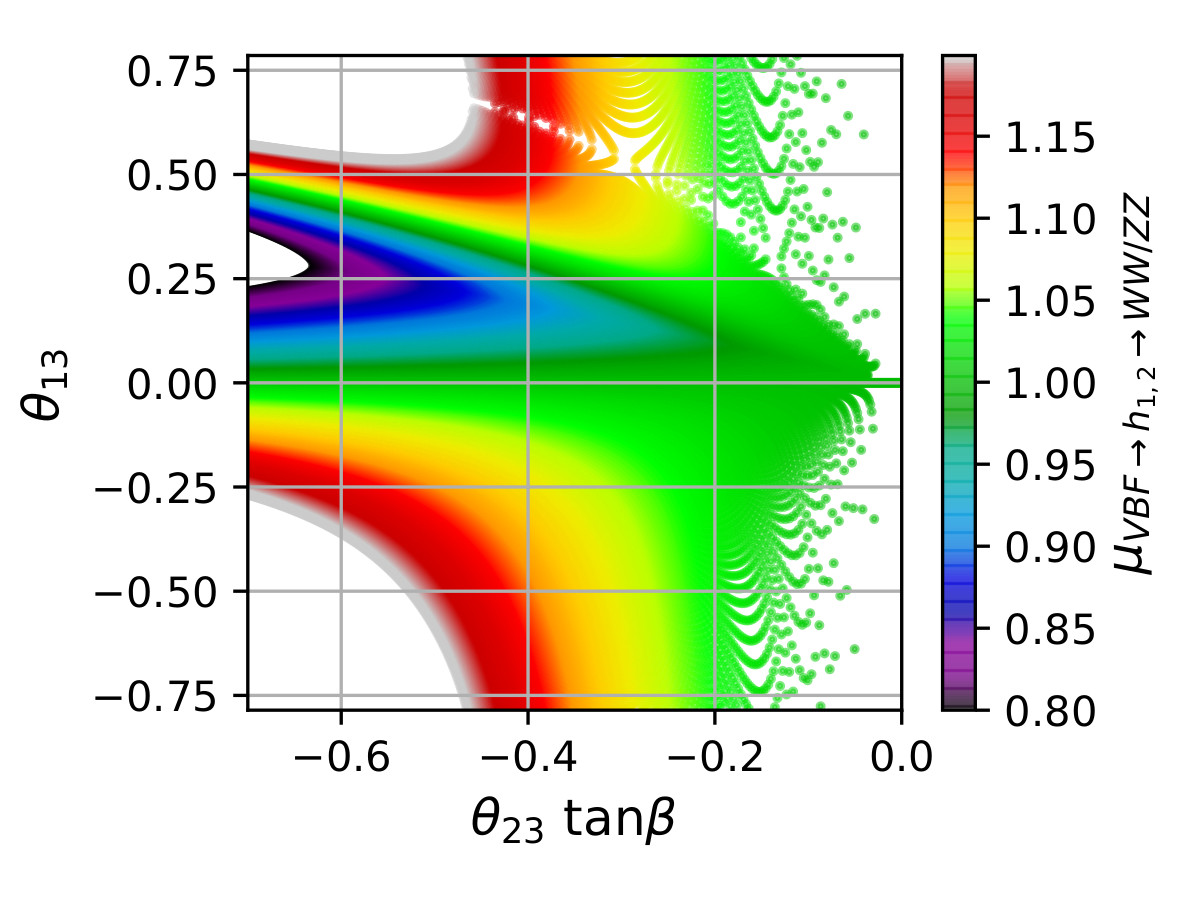}
  \caption{Left panel shows the values of $\theta_{13}$ and
      $\theta_{23}\tan\beta$ for the successful scanned points, the colours
      indicate the value of the signal strength. Right panel show the
      analytic solution for $\theta_{13}$ as a function of
      $\theta_{23}\tan\beta$ and the signal strength described in
      eq.~(\ref{eq:t13th}).}
  \label{fig:test1}
\end{figure}

So far we have seen that the leading behaviour of the signal strengths is
  given by $\theta_{13}$ and $\theta_{23}\tan\beta$. In the limit where
  $\theta_{12}\simeq 0$, we 
  write a biunivocal function to determine one (of these parameters) in terms of the other. %
An approximate relation between $\theta_{13}$ and $\theta_{23}\tan\beta$ might
be useful to study the region around $0.2 \lesssim \theta_{13}\lesssim 0.4$
where it seems possible to mimic the signal of the standard Higgs and make it
indistinguishable even for very precise experimental measurements. %
To determine the relation between the parameters we choose the to
solve the equation:
\begin{eqnarray}
  \label{eq:mudelta}
  \mu_{VBF\rightarrow h_1 \rightarrow WW/ZZ} +
  \mu_{VBF\rightarrow h_2 \rightarrow WW/ZZ} = 1 + \delta.
\end{eqnarray}
By taking $\mu_{VBF\rightarrow h_1 \rightarrow WW/ZZ}$ and
  $\mu_{VBF\rightarrow h_2 \rightarrow WW/ZZ}$ from
  eqs.~(\ref{eq:muan1}) and (\ref{eq:muan2}), neglecting the terms
  proportional $\theta_{12}$, and rewriting the
  $\sin\theta_{13}$ and $\cos(\theta_{13})$ in terms of $\sin(2\,\theta_{13})$
  and $\cos(2\,\theta_{13})$ we can simplify eq.~(\ref{eq:mudelta}) to get a
  quadratic equation in $\cot(2\,\theta_{13})$.
So, there are two solutions for $\theta_{13}$:
\begin{eqnarray}
  \label{eq:t13th}\nonumber
  \cot 2\theta_{13} &=&  \frac{ \mathrm{BR}_{bb}(1+\delta)
  \tan^2\beta\,\theta_{23}^2-\delta}{2\, \delta \, \tan\beta\, \theta_{23}} \\
  && \pm \frac{\sqrt{[\delta+\mathrm{BR}_{bb}(1+\delta) \theta_{23}^2 \tan^2\beta][\mathrm{BR}_{bb}(\mathrm{BR}_{bb}-\overline{\mathrm{BR}}_{bb}\delta)\theta_{23}^2
  \tan^2\beta-\overline{\mathrm{BR}}_{bb}\delta]}}{2 \sqrt{\mathrm{BR}_{bb}}\,
     \delta\, |\tan\beta \, \theta_{23}|}
\end{eqnarray}
where $\overline{\mathrm{BR}}_{bb}= 1-\mathrm{BR}_{bb}$.
For $\delta=0$ the solution simplifies to 
\begin{eqnarray}
  \label{eq:t13th0}
  \cot 2\theta_{13}=\frac{1+\mathrm{BR}_{bb}(-4+\theta_{23}^2\,\tan^2\beta)}{4\mathrm{BR}_{bb}\,\theta_{23}\,\tan\beta}
\end{eqnarray}
With eq.~(\ref{eq:t13th}) we are able to determine $\theta_{13}$ in terms of
$\theta_{23}\tan\beta$ and $\delta$.
Figure \ref{fig:test1} shows the comparison between the semi-analytical relation in
eq.~(\ref{eq:t13th}) and the numerical results from our scans. Although it is not a precise relation,
eq.~(\ref{eq:t13th}) gives a very good approximation to the correlation between 
$\theta_{13}$ and $\theta_{23}$ for a fixed value of $\delta$. 

\section{Searching for mass-degenerate Higgses}
\label{sec:signal}

As commented in references \cite{Gunion:2012he} and \cite{Grossman:2013pt}
there are ways to test the existence of mass-degenerate states. The
determinant of a signal strengths square matrix could give information
about the number of resonances. If the determinant of the square matrix is equal to zero
then the existence of a single Higgs resonance will be enough to reproduce the
signal strengths.\\
For simplicity we will use a compact notation:
  $\mu_{ij}=\mu_{i\rightarrow j}$, where $i$ represents the production mode 
  and $j$ the decay channel. Considering two square matrices,%
\begin{eqnarray}
  \label{eq:Rmat}
  R^A=\left( \begin{array}[h]{cc}
               \mu_{GGF,\gamma\gamma} & \mu_{GGF,\tau\tau}\\
               \mu_{VBF,\gamma\gamma} & \mu_{VBF,\tau\tau}
             \end{array} \right), \ \ %
  R^B=\left( \begin{array}[h]{cc}
               \mu_{GGF,\gamma\gamma} & \mu_{GGF,WW}\\
               \mu_{VBF,\gamma\gamma} & \mu_{VBF,WW}
             \end{array} \right)                                  
\end{eqnarray}
the condition for the determinant to be non-zero can be written in
terms of the ratios
\begin{eqnarray}
  \label{eq:rs}
  \frac{\mu_{VBF,WW}}{\mu_{VBF,\gamma\gamma}}\neq
  \frac{\mu_{GGF,WW}}{\mu_{GGF,\gamma\gamma}} \ \ \ \mathrm{and} \ \ \ 
  \frac{\mu_{VBF,\tau\tau}}{\mu_{VBF,\gamma\gamma}}\neq
  \frac{\mu_{GGF,\tau\tau}}{\mu_{GGF,\gamma\gamma}}. 
\end{eqnarray}
\begin{table}[h]
  \centering
  \setlength{\extrarowheight}{3pt}
  \begin{tabular}[h]{|c|c|}
    \hline
    Parameter & ATLAS + CMS\\
    \hline
    $\mu_{V,\gamma\gamma}$ & $1.05^{+0.44}_{-0.41}$\\
    $\mu_{V,ZZ}$ & $0.47^{+1.37}_{-0.92}$\\
    $\mu_{V,WW}$ & $1.38^{+0.41}_{-0.37}$\\
    $\mu_{V,\tau\tau}$ & $1.12^{+0.37}_{-0.35}$\\
    $\mu_{V,bb}$ & $0.65^{+0.31}_{-0.29}$\\
    $\mu_{F,\gamma\gamma}$ & $1.16^{+0.27}_{-0.24}$\\
    $\mu_{F,ZZ}$ & $1.42^{+0.37}_{-0.33}$\\
    $\mu_{F,WW}$ & $0.98^{+0.22}_{-0.20}$\\
    $\mu_{F,\tau\tau}$ & $1.06^{+0.60}_{-0.56}$\\
    $\mu_{F,bb}$ & $1.15^{+0.99}_{-0.94}$\\
    \hline
  \end{tabular}
  \caption{Ten parameter fit of $\mu_F^f$ and $\mu_{V}^f$. Table 15 of
    reference \cite{Khachatryan:2016vau}}
  \label{tab:muexp}
\end{table}
To check if it is possible to establish the existence of two resonances in the
NMSSM we consider the set of pNMSSM posterior sample 
described in section~\ref{sec:scan} and check for points which are within one and three sigma of the
particular signal strengths listed in Table~\ref{tab:muexp}.

\begin{figure}[t!]
  \centering
  \includegraphics[width=0.45\textwidth]{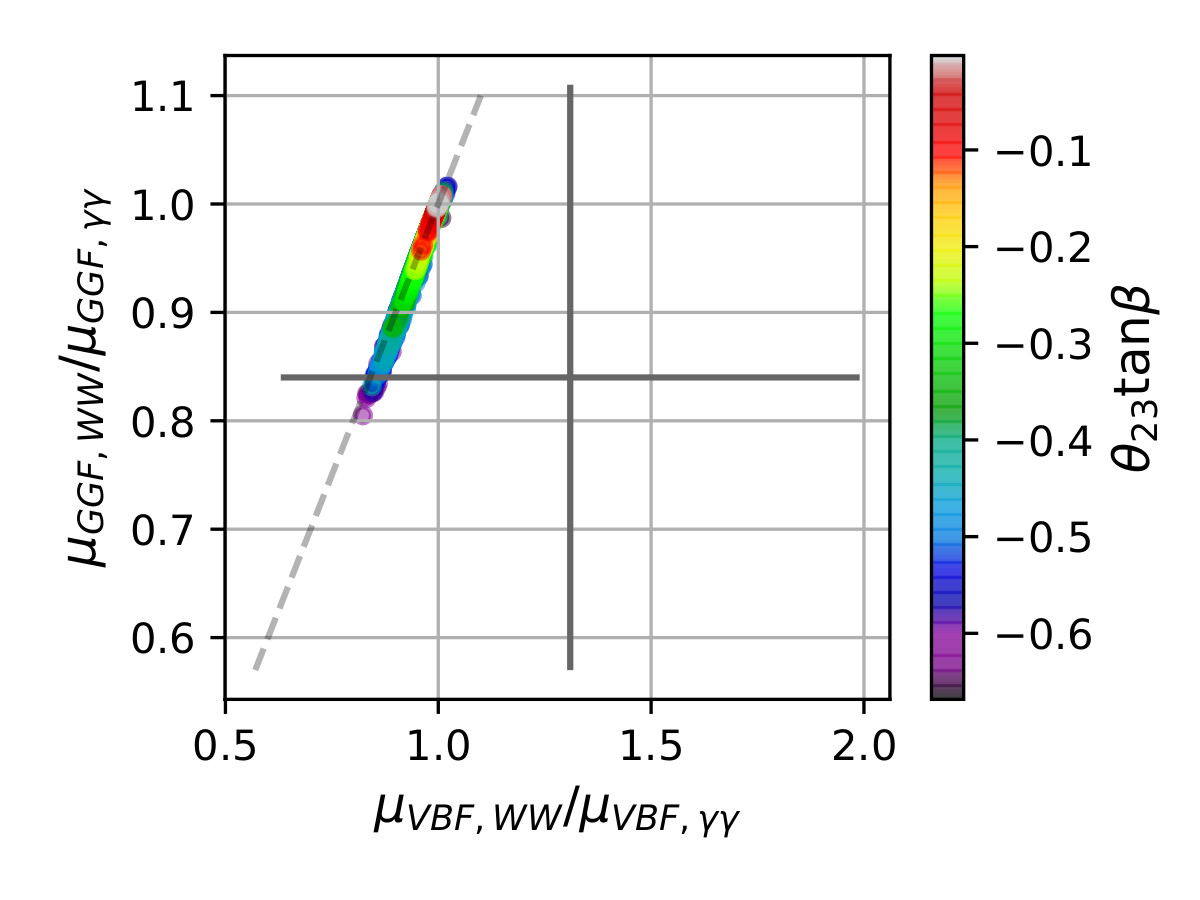} %
  \includegraphics[width=0.45\textwidth]{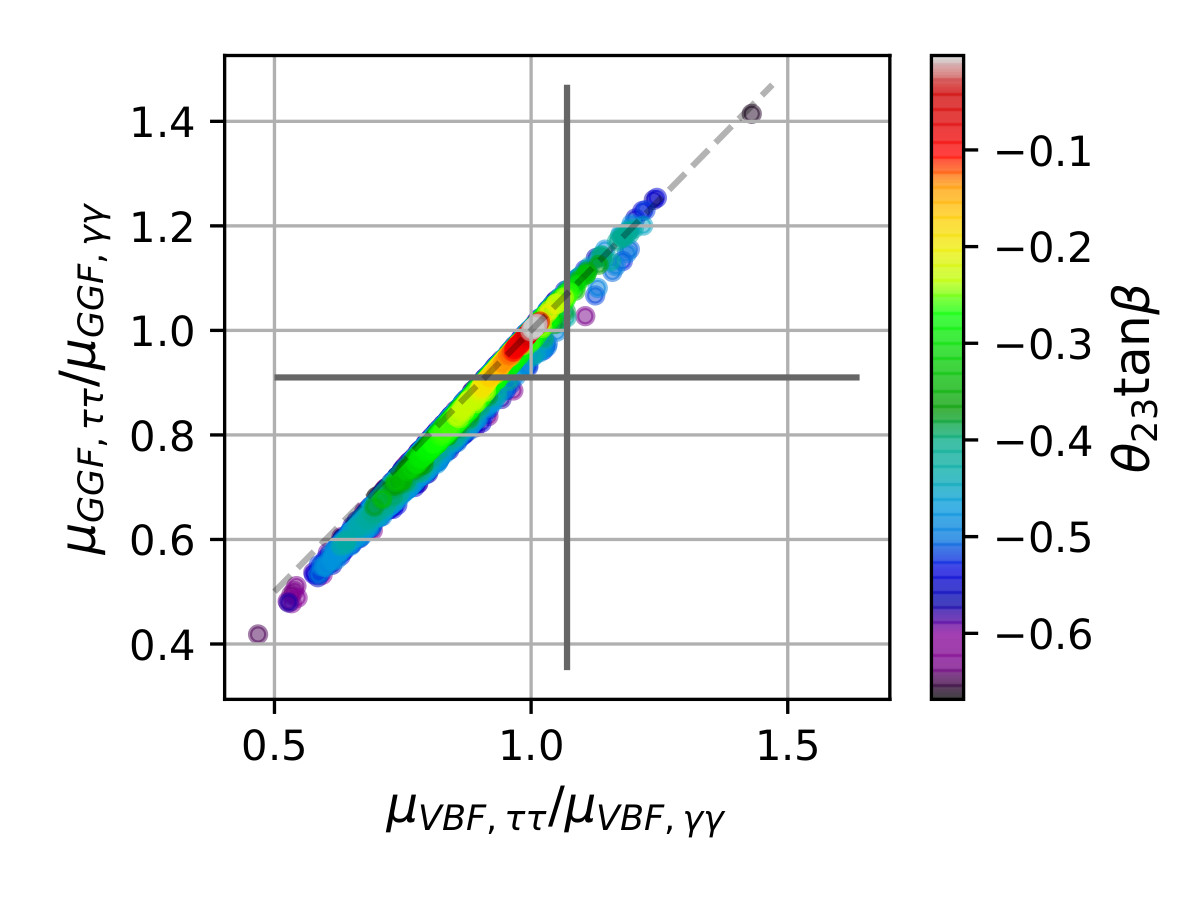}\\%
    \includegraphics[width=0.45\textwidth]{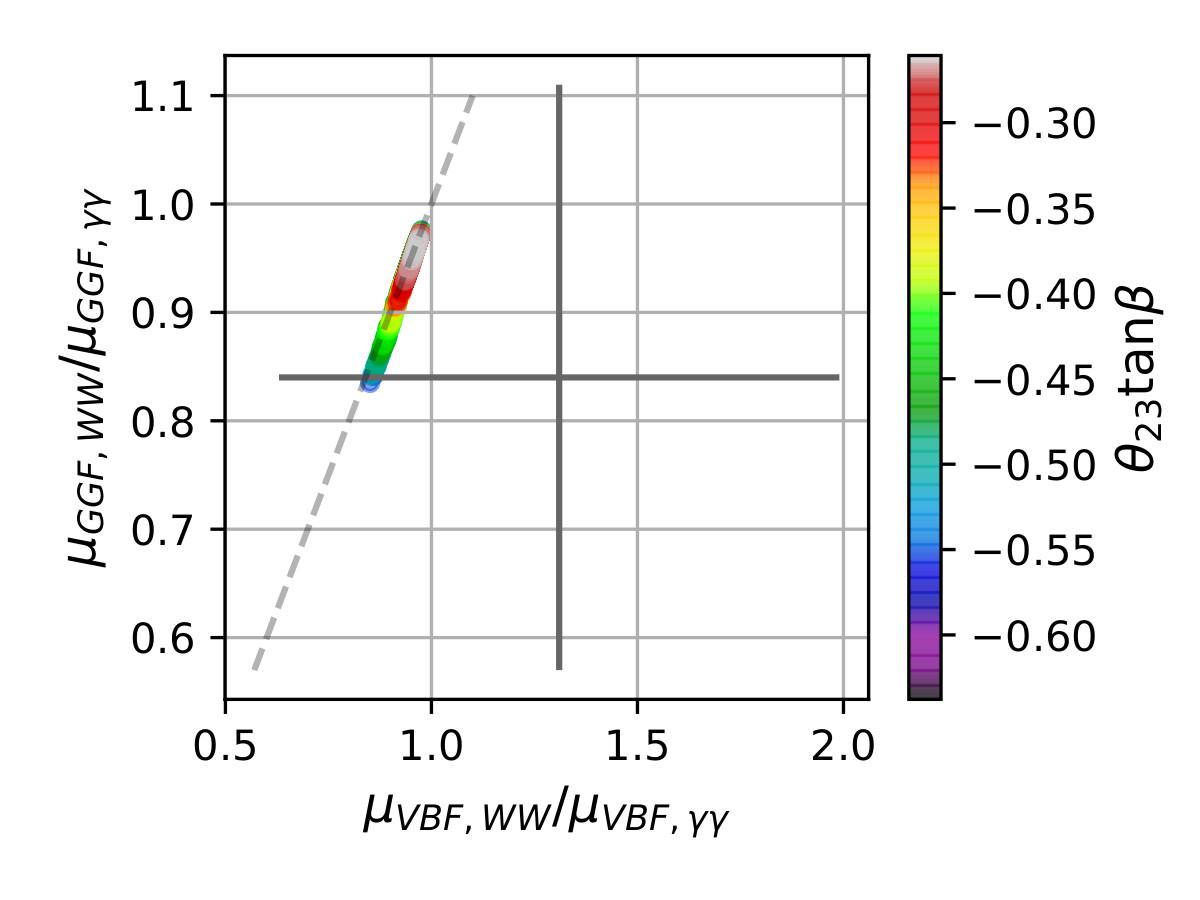} %
  \includegraphics[width=0.45\textwidth]{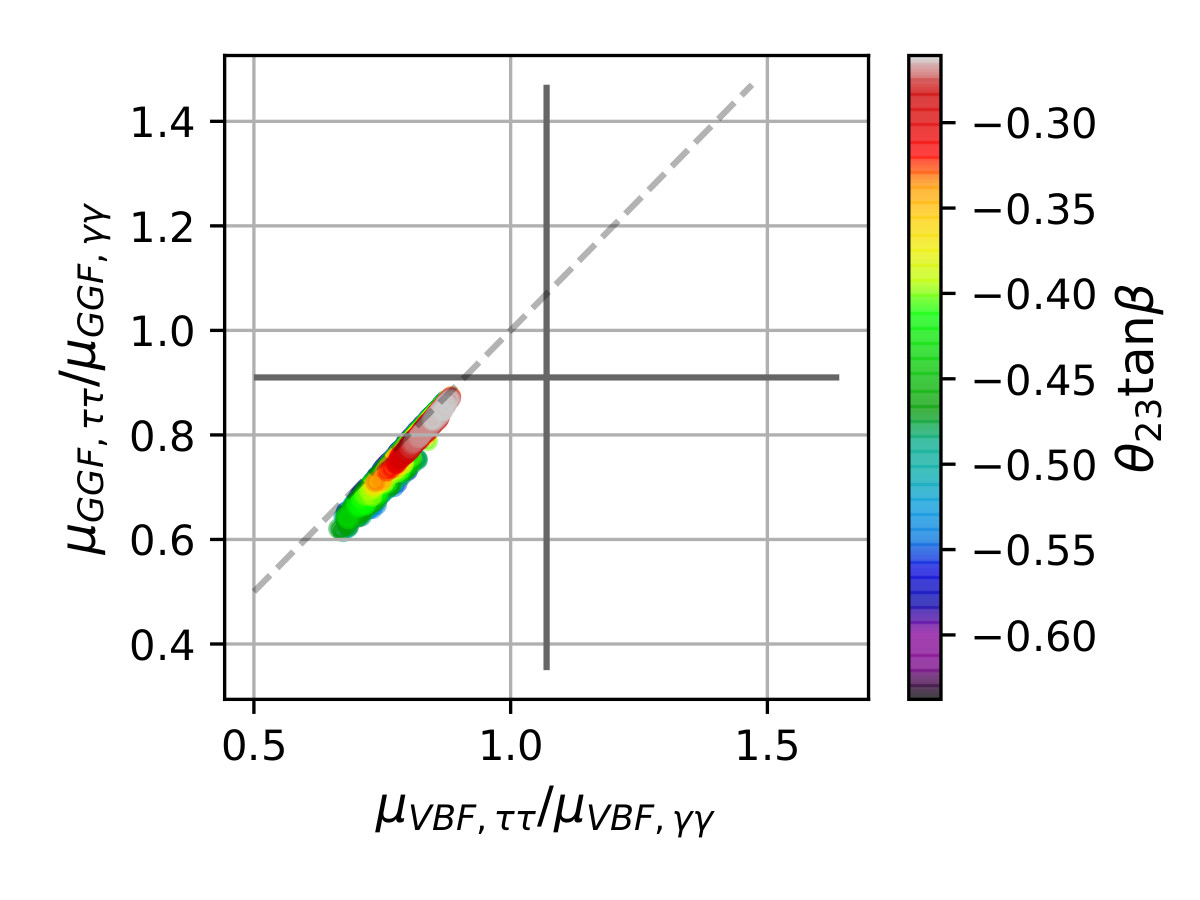}
  \caption{Comparison between ratios of the signal strengths from the pNMSSM sample considered.
    The Upper (lower) row shows points with individual signal strength within three (one) sigma
    with respect to the experimental values. The large dark gray lines
      represent the error bars of the experimental value for each of the
      rates. The dotted line indicates det($R$)=0.}
  \label{fig:ratios}
\end{figure}

Figure~\ref{fig:ratios} shows the comparison between the ratios of the
signal strengths in eq.~(\ref{eq:rs}). The upper (lower) panel shows all the
points that are within three (one) sigma of the values of the individual
rates. The points are ordered in such a way that smaller values of
$|\theta_{23}\tan\beta|$ are on top. Notice that in the lower panel the one
sigma region do not contain the point $\{1,1\}$, which is what we expect from
a standard Higgs, this is because the experimental value of $\mu_{VBF,bb}$ is
$0.65^{+0.31}_{-0.29}$ (see Table~\ref{tab:muexp}), it doesn't include the SM
value at one sigma. The left panel of Figure~\ref{fig:ratios} shows that the
ratios between $WW$ and $\gamma\gamma$ signal strength are basically the same,
meaning that the determinant of $R_A$ is approximately zero and therefore in
agreement with a single resonance hypothesis. On the other hand the ratios
between $\tau\tau$ and $\gamma\gamma$ signal strength are slightly separated
from the dotted line, the determinant of $R_B$ is different from
zero. In general we would expect that if there is more than one Higgs
  state the ratio between two signal strengths with the same production
  process and different decay product is not going to be equal to
  one. However, we get that this ratio is almost the same for the rate between
  gluon-gluon fusion and for vector-boson fusion production processes, which
  indicates that both production cross-sections are very similar for a given
  Higgs state. Therefore, it doesn't seem possible to distinguish between
  single and double resonances from those measurements for this set of scanned
  points. \\ 
%

Is there any observable that could be used to distinguish between single and double resonance
signals? From the discussion of the previous sections we have learned
  that $\mu_{VBF,bb}$ have an opposite behaviour with respect to the other signal
  strength we have considered, therefore we may suspect
  that the production of Higgs states associated to bottom quarks
compared to the production associated to vector bosons would give a larger
departure from the SM signal than the comparison
between vector-boson fusion and gluon-gluon fusion.\\
Let us consider the matrices,%
\begin{eqnarray}
  \label{eq:Rmat2}
  R^C=\left( \begin{array}[h]{cc}
               \mu_{BBF,\gamma\gamma} & \mu_{BBF,\tau\tau}\\
               \mu_{VBF,\gamma\gamma} & \mu_{VBF,\tau\tau}
             \end{array} \right), \ \ %
  R^D=\left( \begin{array}[h]{cc}
               \mu_{BBF,\gamma\gamma} & \mu_{BBF,WW}\\
               \mu_{VBF,\gamma\gamma} & \mu_{VBF,WW}
             \end{array} \right)                                  
\end{eqnarray}
where BBF represents the Higgs productions associated to bottom
quarks. To obtain a determinant different from zero requires that
  ratios of the signal strengths follow:
\begin{eqnarray}
  \label{eq:rs2}
  \frac{\mu_{VBF,WW}}{\mu_{VBF,\gamma\gamma}}\neq
  \frac{\mu_{BBH,WW}}{\mu_{BBH,\gamma\gamma}} \ \ \ \mathrm{and} \ \ \ 
  \frac{\mu_{VBF,\tau\tau}}{\mu_{VBF,\gamma\gamma}}\neq
  \frac{\mu_{BBH,\tau\tau}}{\mu_{BBH,\gamma\gamma}}
\end{eqnarray}
To compute the signal strength of Higgs production associated to bottom quarks we
use the reduced couplings to bottom quarks computed by NMSSMtools.
\begin{figure}[t!]
  \centering
  \includegraphics[width=0.45\textwidth]{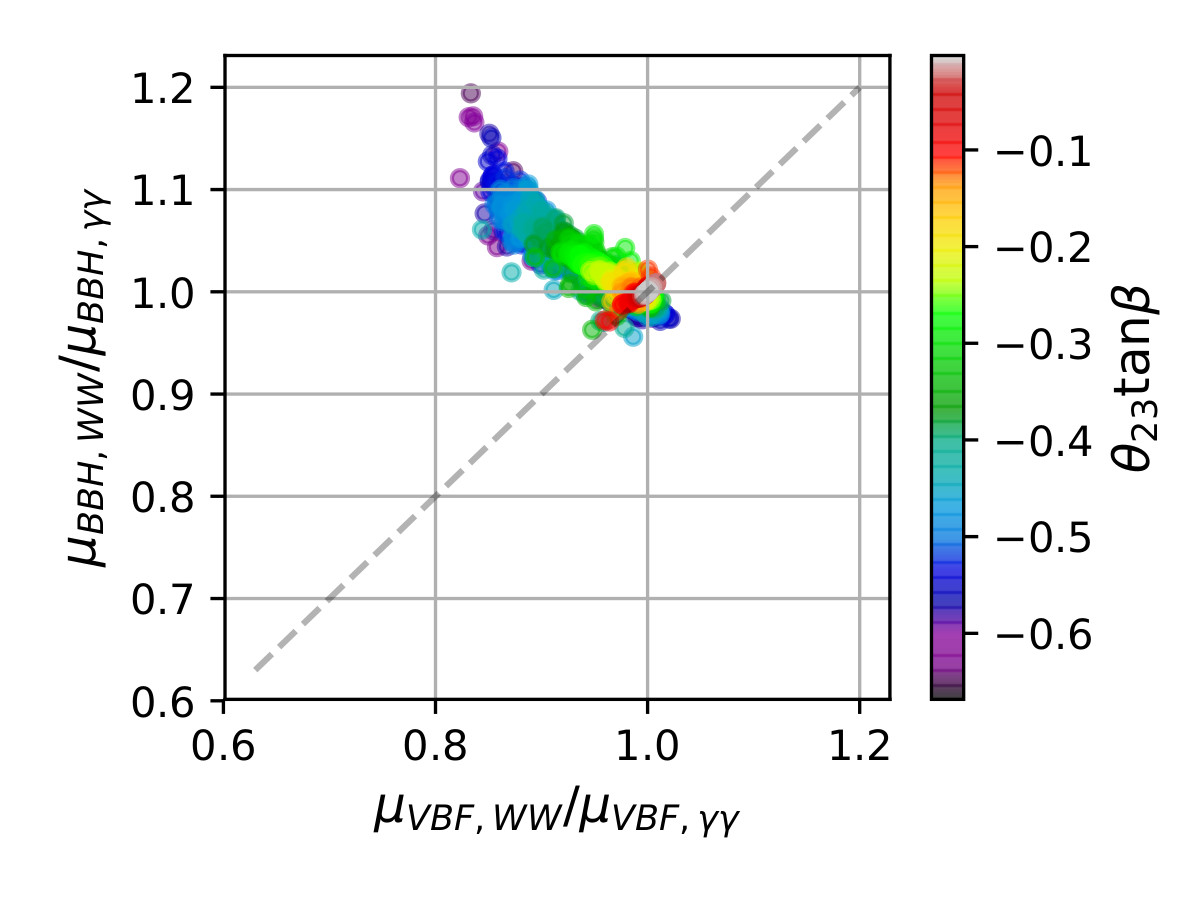} %
  \includegraphics[width=0.45\textwidth]{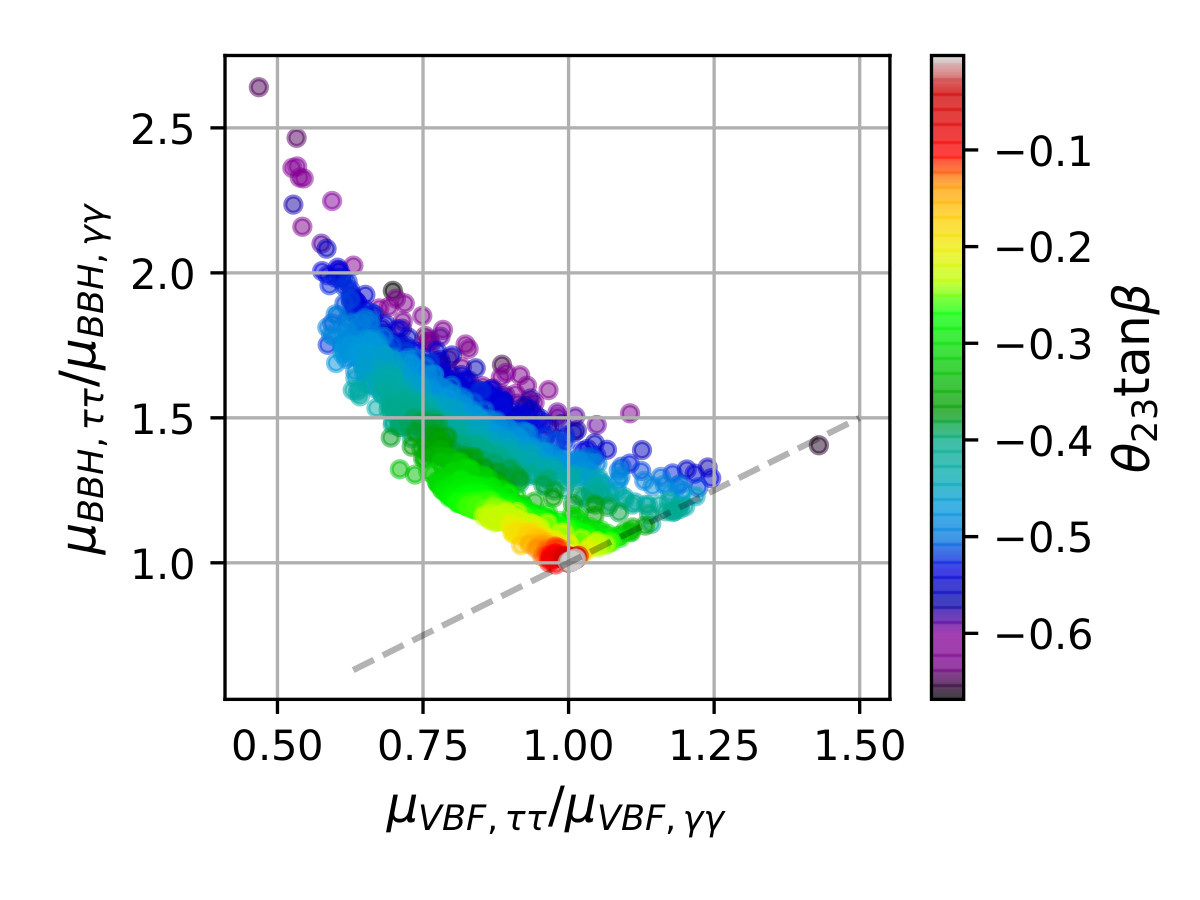}
  \caption{Comparison between vector-boson fusion and Higgs production
    associated to bottom quarks. Points are within three sigma of the
    measured individual signal strengths. The dotted line indicates
    det($R$)=0.}
  \label{fig:ratios2}
\end{figure}
Figure~\ref{fig:ratios2} shows the comparison of the ratios described in
eq.~(\ref{fig:ratios2}) for points that fulfill the experimental signal
strength listed in Table~\ref{tab:muexp} within three sigma. The figure shows
that the determinant of the $R^C$ and $R^D$ is different from zero for a large
part of the points, and therefore it gives a clear signature for the existence
of more than one Higgs resonance.\\

It may be surprising to see such a large deviation from zero in the
determinant of $R^C$ and $R^D$ and not in the determinant of $R^A$ and $R^B$,
the main reason lies in the difference between the production
processes. Although it does not seem straight forward from the analytic
expressions of the full signal strength to single out this differences and
directly relate them with the value of the determinants, one can always
compare the production cross-sections for each Higgs state separately. If they
are approximately the same, then the ratios shown in Figures~\ref{fig:ratios} and
\ref{fig:ratios2} will be the same --  and the 
determinant of the matrix $R$ will be approximately equal to zero.\\

For simplicity let us consider that the gluon-gluon fusion cross section is
dominated by the coupling of the Higgs to top quarks, this consideration will
allow us to have more insights of the source of discrepancy between
the determinants. Eqs.(\ref{eq:couplings}) show that $\hat{g}_{tth_i}$ has an
extra factor $-U_{i2}/\tan\beta$ with respect to the coupling to vector
bosons, using the approximation of small $\theta_{23}$ and negligible
$\theta_{12}$, the extra factor simplify to $\theta_{23}/\tan\beta$ times
$\cos\theta_{13}$($\sin\theta_{13}$) for $h_1$($h_2$), a factor suppressed by
$\tan\beta$. Therefore, unless $\tan\beta$ is close to one, or $\theta_{23}$
is large, we would expect very similar signal strengths for gluon-gluon fusion
and vector-boson fusion for each Higgs state, in consequence the total signal
strengths for the same final state will be also very similar, and the
determinant of $R^A$ and $R^B$ will be close to zero.\\

On the contrary, if instead of gluon-gluon fusion production process we
consider Higgs production associated to bottom quarks,
eqs.~(\ref{eq:couplings}) show that $\hat{g}_{bbh_i}$ has an extra factor
$U_{i2}\tan\beta$ with respect to vector boson coupling, the factor is
$\tan^2\beta$ larger than in the case of $\hat{g}_{tth_i}$. For non-negligible
values of $\theta_{23}$ there will be a significant departure of signal
strength of the Higgs production associated to bottom quarks with respect to the
vector-boson fusion for the same final state. When computing the ratio of the
total signal strength for different final states we would expect a larger
deviation, in consequence the determinant of $R^C$ and $R^D$
will be different from zero.\\

\begin{figure}[t!]
  \centering
  \includegraphics[width=0.45\textwidth]{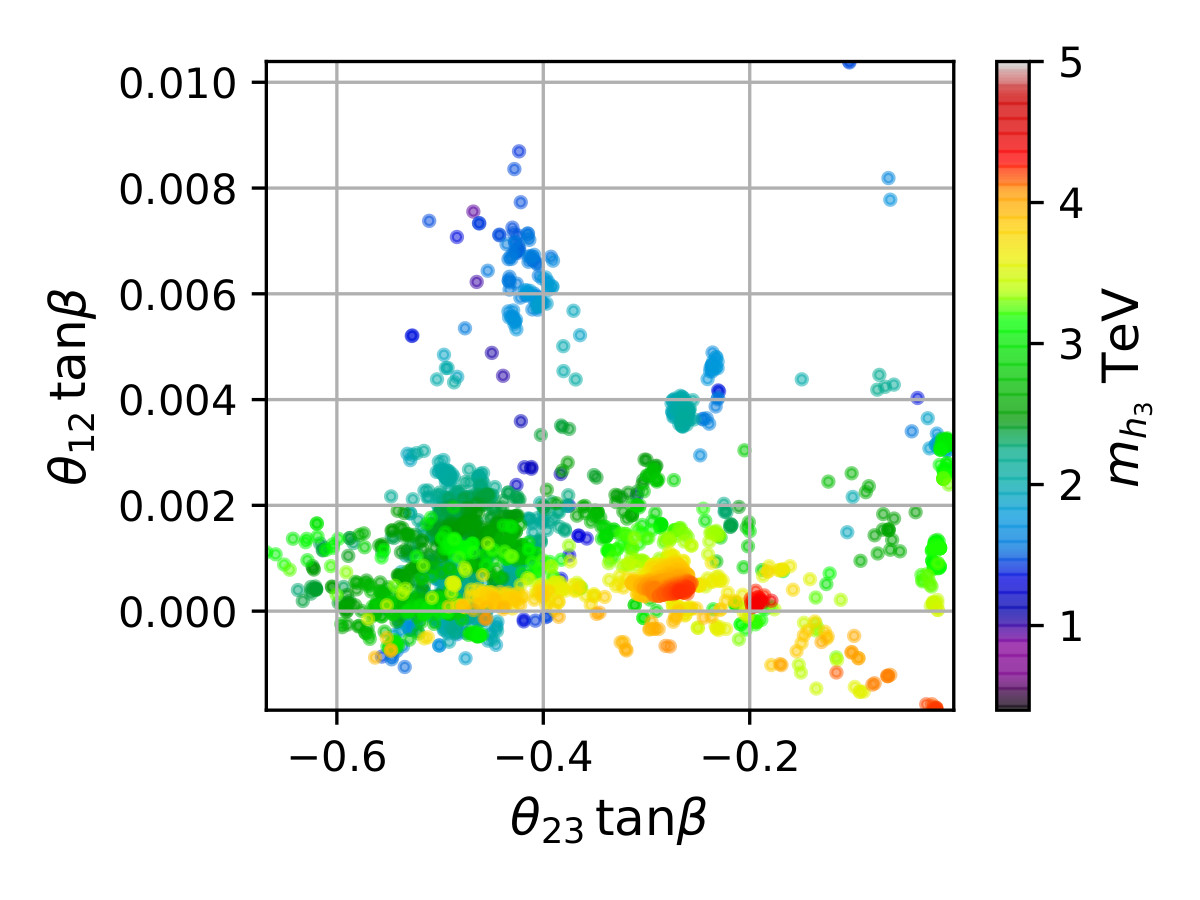} \ \ \ %
  \includegraphics[width=0.45\textwidth]{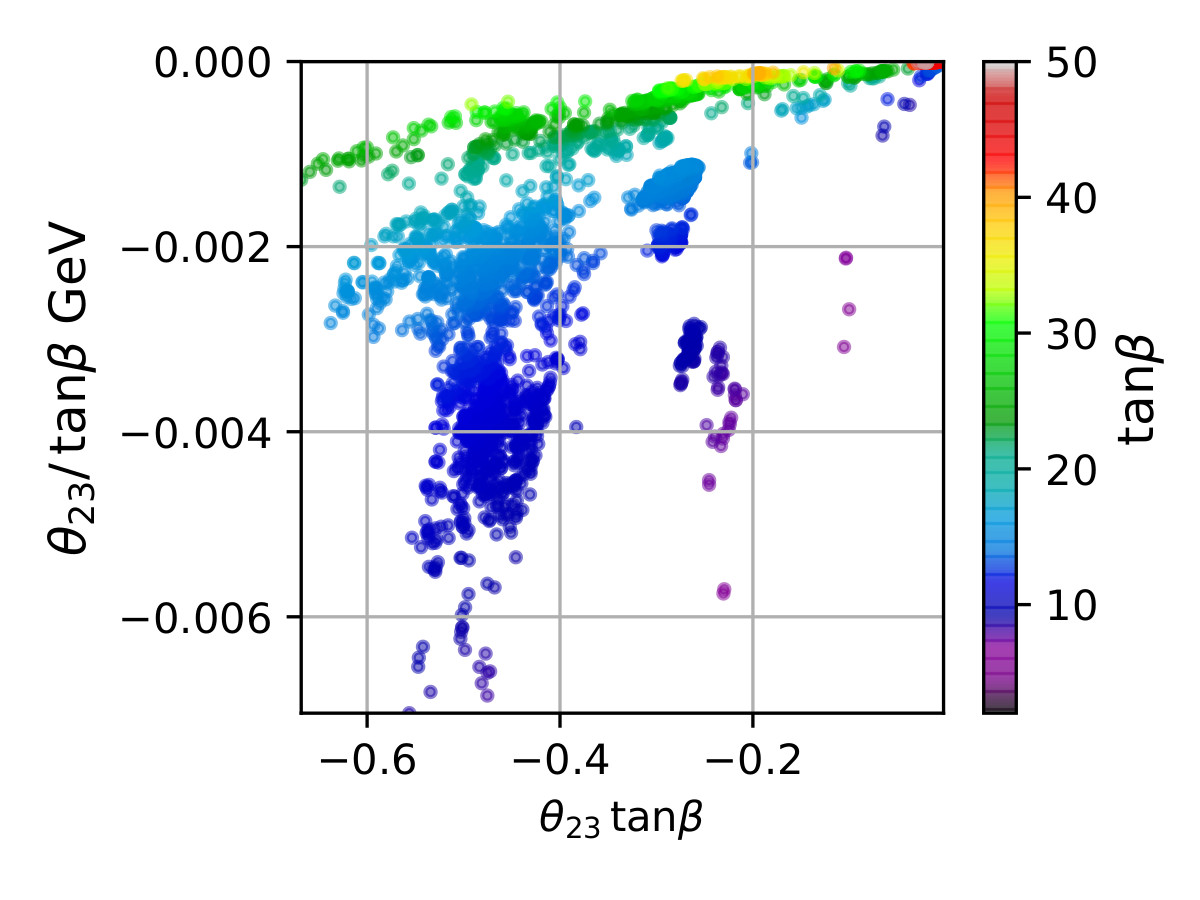}
  \caption{Left panel shows the values of $\theta_{12}$, $\theta_{23}$ and the
    mass of the heaviest CP-even Higgs in the colour bar. Right panel
    illustrate the $\theta_{23}$ dependence in the coupling of the Higgs for
    up-type quarks ($\theta_{23}/\tan\beta$) and down-type quarks and leptons
    ($\theta_{23}\tan\beta$).}
  \label{fig:t12_t23_tanb}
\end{figure}

These arguments describe very well a set of points with medium to large values
of $\tan\beta$. For small values of $\tan\beta$ and large enough values of
$\theta_{23}$ the determinant of $R^A$ and $R^B$ will also show a departure
from unity. Figure~\ref{fig:t12_t23_tanb} shows the values of $\theta_{12}$,
$\theta_{23}$, $\tan\beta$ and $m_{h_3}$ for the pNMSSM posterior sample 
 with $m_{h_3}$ larger that 1 TeV and values of 
$\tan\beta$ larger than 10. As we expected the value of
$\theta_{23}/\tan\beta$ is tiny, which explains why the determinant of $R^A$
and $R^B$ is very close to zero. The large values of $\tan\beta$ also explain
the large departure from one for the determinant of $R^C$ and
$R^D$.\\

\begin{figure}[t!]
  \centering
  \includegraphics[width=0.45\textwidth]{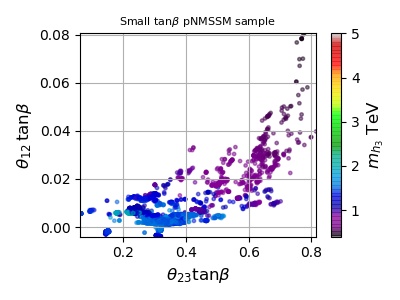} \ \ \ %
  \includegraphics[width=0.45\textwidth]{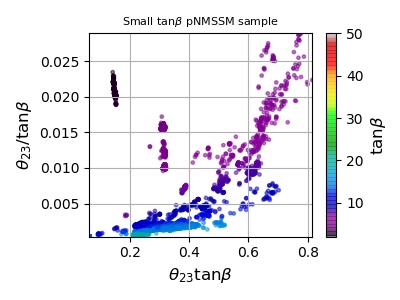} \\
  \includegraphics[width=0.45\textwidth]{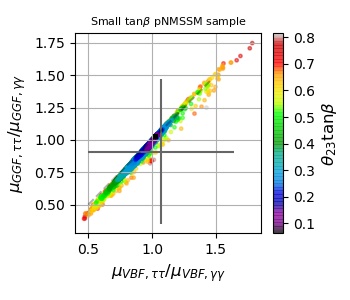} \ \ \ %
  \includegraphics[width=0.45\textwidth]{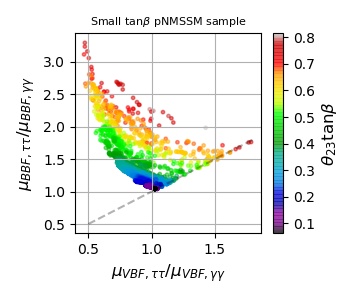}
  \caption{Top row shows the values of $\theta_{12}$ and $\theta_{23}$
    with respect to $\tan\beta$ and $m_{h_3}$. Bottom panel show the comparison
    between the signal strengths of  vector-boson fusion and gluon-gluon
    fusion (left), vector-boson fusion and Higgs production associated to
    bottom quarks (right), for $\tau\tau$ and $\gamma\gamma$ final states.}
  \label{fig:chain5}
\end{figure}

 Our scan focused on the region of the parameter space with medium to large
values of $\tan\beta$, to complete our analysis we analyse a new set of points
with smaller values of $\tan\beta$ relative to the first sample set. We perform another small scan giving more
preference to the region of small $\tan\beta$ and small $m_{h_3}$,
covering $\tan\beta$ in the range of $[2.5,21]$ and $m_{h_3}$ in the
  range of [435 GeV to 2 TeV], the results are summarized in
figure~\ref{fig:chain5}.  The top row of the figure shows the values of
$\theta_{12}$ and $\theta_{23}$ with respect to $m_{h_3}$ and $\tan\beta$. To analyse
these two plots in comparison with Figure~\ref{fig:t12_t23_tanb} we have used
the same range for the variables plotted in the colour bar to make easier the
comparison. First let us focus on the top-left plot of Figure~\ref{fig:chain5}.
Note that the range of values for
$|\theta_{23}\tan\beta|$ is almost the same for both samples suggesting that
this parameter is directly constrained by the experimental measurements of the
Higgs couplings. Smaller values of $m_{h_3}$ are correlated with larger values
of $\theta_{12}$, still $|\theta_{23}|$ is one order of magnitude larger than
$|\theta_{12}|$, meaning that the approximation of $\theta_{12}\sim 0$ is
still valid. The top-right plot of Figures~\ref{fig:t12_t23_tanb} and
\ref{fig:chain5} compare the values of $\theta_{23}\tan\beta$ with
$\theta_{23}/\tan\beta$ that illustrate the contribution of $\theta_{23}$ to
the Higgs production associated to bottom quarks (x-axis) and gluon-gluon
fusion production (y-axis).\\

The bottom row of Figure~\ref{fig:chain5} show the values of $R^B$ and $R^D$
for the new set of scanned points. Here, points with
$\theta_{23}\tan\beta \sim 0.7$ correspond to $|\theta_{23}/\tan\beta|$ up to
0.030, which is around fifty times larger than our first scan. This increment
will be reflected in the value of $R^B$, which involves the rate plotted in
the left panel of the figure. Previous studies, like \cite{Gunion:2012he,
  Munir:2013wka, Moretti:2015bua} pointed out that the determinant of $R^A$
and $R^B$ will be useful to determine the existence of more than one
resonance. Our analyses indicate that this is indeed the case but
  mostly for pNMSSM regions with relatively smaller $\tan\beta$ values and
  lighter $h_3$.
The botton-right plot of Figure~\ref{fig:chain5} shows the relevant ratios to
compute the determinant of $R^D$.  There is a discrepancy in the region with
$|\theta_{23}\tan\beta|$ larger than $\sim 0.65$. According to the top-row
plots of Figure~\ref{fig:chain5}, points with $|\theta_{23}\tan\beta|>0.7$
correspond to $m_{h_3}$ smaller that 1 TeV and $\tan\beta$ smaller than
10. Getting relatively larger values for $|\theta_{23}\tan\beta|$ in the new
set of points scanned compared to the first pNMSSM posterior sample is in
accord with the fact that $|\theta_{23}|$ increases as $m_{h_3}$ decreases for
a fixed value of $\lambda$ (as discussed in section~\ref{higgsconsts}). So in
the new scan by exploring $m_{h_3} < $ 1 TeV, we expand the range of
exploration for $|\theta_{23}\tan\beta|$.

\section{Conclusions}

We studied the phenomenology of the two mass degenerate CP-even Higgs bosons
in the NMSSM using a sample set from the parameter scan of the pNMSSM. In this
scenario it is possible to reproduce the experimental signal measured by ATLAS
and CMS. We parameterised the Higgs boson signal strengths using three
  angles and found that it is possible to write approximate expressions in
  terms of two parameters $\theta_{23}\tan\beta$ and
  $\theta_{13}$, where $\theta_{23}$ is the mixing between the singlet
  and the heaviest neutral Higgs of the Higgs doublet $H_0$ and $\theta_{13}$
  the mixing between the lightest neutral scalar of the Higgs doublet and the singlet. 
We have focused our
analysis into observables that could help to determine the existence of more
that one Higgs state, leading to the following conclusions. 
\begin{itemize}
\item To obtain two mass degenerate CP-even Higgs bosons there is required
  tuning associated to large values of $A_\kappa$, $\lambda$,
  $\kappa$ and $\mu$. An approximate relation between those parameters could
  be obtained from the tree level mass relations, although this relation
  simplifies the expression for the mass of the lightest pseudoscalar it does
  not point out to specific mass relations.
  
\item An approximate expression for $\theta_{23}$ can be written in terms of
  $\mu/\lambda$ and $\tan\beta$. The allowed range for
  $|\theta_{23}\tan\beta|$ is between 0.0 and 0.7. Greater values can
  be obtained if $m_{h_s} \lesssim 1$ TeV and $\tan \beta \lesssim 8$ are imposed.
  There 
  are no direct constraints on the mass spectra from specific values of
  $\theta_{23}$ but 
  it is 
  possible to reproduce various values of 
  $m_{h_3}$ for a fixed value of $\theta_{23}$ and different values of
  $\lambda$.

\item Analysing the Higgs bosons couplings to fermions and vector bosons, and
  the signal strengths, we found that the signal of the superposition of
    the Higgs bosons decaying to leptons (and bottom quarks) depart from the SM
  signal in an opposite direction with respect to vector boson final
  states. This is 
  proportional to $|\theta_{23}\tan\beta|$. 

\item With respect to expectations due to previous studies, it was surprising to find 
  that for medium to large values of $\tan\beta$, it is rather difficult to distinguish
  the two degenerate Higgs from the single Higgs scenario when the matrix of
  signal strengths are for vector-boson and gluon-gluon fusion Higgs productions (with
  the Higgs decaying to vector boson).

\item By including Higgs production in association with bottom quarks in the
  signal strengths square matrix we found that the matrix determinant departs significantly
  large from the single resonance value.
  Therefore the process $p p \rightarrow bbh$ can be an important channel in searches for multiple
  Higgs states degenerate around $125 \GeV$. 

\end{itemize}

\section*{Acknowledgment}

Thanks to Alberto Casas for very useful comments and discussions, and to Fernando Quevedo for
encouragements towards the NMSSM project. Maria Cabrera thanks ICTP and CERN Theory Division
for hosting and supporting her as short-term visitor.

\bibliography{references}{}
\bibliographystyle{hieeetr}

\end{document}